\documentclass[aps,prx,amsmath,amssymb,onecolumn,superscriptaddress,showpacs,floatfix,notitlepage]{revtex4-2}

\makeatletter
\renewcommand{\p@subsection}{}
\renewcommand{\p@subsubsection}{}
\makeatother

\usepackage{bm}
\usepackage{epsfig}
\usepackage[usenames]{color}
\usepackage{soul}
\usepackage{graphicx, xcolor}
\usepackage{wrapfig}
\usepackage{subcaption}
\usepackage{hyperref}
\usepackage{amsmath}
\usepackage{mathtools}

\graphicspath{ {./images/} }

\begin{document}


\title{Nonreciprocal total cross section of quantum metasurfaces}

\author{Nikita~Nefedkin}
\affiliation{Photonics Initiative, Advanced Science Research Center, City University of New York, NY 10031, USA}
\email{aalu@gc.cuny.edu} \email{nnefedkin@gc.cuny.edu}

\author{Michele~Cotrufo}
\affiliation{Photonics Initiative, Advanced Science Research Center, City University of New York, NY 10031, USA}

\author{Andrea~Al\`{u}}
\affiliation{Photonics Initiative, Advanced Science Research Center, City University of New York, NY 10031, USA}
\affiliation{Physics Program, Graduate Center, City University of New York New York, NY 10016, USA}

\date{\today}

\begin{abstract}

  Nonreciprocity originating from classical interactions among nonlinear scatterers has been attracting increasing attention in the quantum community, offering a promising tool to control excitation transfer for quantum information processing and quantum computing. In this work, we explore the possibility of realizing largely nonreciprocal total cross sections for a pair of quantum metasurfaces formed by two parallel periodic arrays of two-level atoms.
  We show that large nonreciprocal responses can be obtained in such nonlinear systems by controlling the position of the atoms and their transition frequencies, without requiring that the environment in which the atoms are placed is nonreciprocal.
  We demonstrate the connection of this effect with the population of a slowly-decaying dark state, which is critical to obtain large nonreciprocal responses.

\end{abstract}

\maketitle


\section{Introduction}\label{sec:intro}

Combining the classical concepts of metasurfaces and metamaterials with quantum effects has attracted increasing attention over the past few years~\cite{bekenstein_quantum_2020,solntsev2021metasurfaces,wang2018quantum,sakhdari2016efficient,chen2015modulatable}.
Recently, ensembles of cold atoms in free space have emerged as a novel platform to control light-matter interactions at the few-photon level.
In these systems, often termed \textit{quantum metamaterials}, exotic phenomena emerge due to interplay between the field radiated and scattered by multiple atoms~\cite{reitz_cooperative_2022,ballantine_quantum_2021, masson_many-body_2020,ballantine_unidirectional_2022, bekenstein_quantum_2020, fernandez2022tunable}.
Moreover, these systems offer exciting opportunities for applications in quantum information processing and metrology in free space~\cite{guimond_subradiant_2019,chang_colloquium_2018, reitz_cooperative_2022, manzoni2018optimization}.

Among the many desirable features for quantum and classical computation, there is a strong need for devices that support nonreciprocal wave propagation, i.e., systems in which light is transmitted along opposite directions with different efficiencies. Several approaches for nonreciprocity have been recently suggested in both the classical~\cite{caloz2018electromagnetic,nassar2020nonreciprocity,manipatruni2009optical} and quantum realms~\cite{mahoney2017chip,zhang2018thermal,hamann2018nonreciprocity}. In particular, in recent works it has been shown that nonreciprocal propagation or scattering can be obtained in nonlinear systems if they couple asymmetrically to different input/output channels. In classical electromagnetism, this approach has been demonstrated with resonators (such as optical cavities or radio-frequency circuits) coupled asymmetrically to two ports and loaded with Kerr-like nonlinearities, such that their resonant frequencies depend on the stored energy \cite{sounas_fundamental_2018,sounas2018broadband,yang_inverse-designed_2020,cotrufo2021nonlinearityPart1,cotrufo2021nonlinearityPart2}. As a consequence, the transmission can be markedly different along the two directions. In parallel, similar behaviors have been shown in quantum devices by suitably combining quantum nonlinearities with breaking of spatial symmetry \cite{Roy2010Fewphoton, Roy2013Cascaded, Fratini2014FabryPerot,Fratini2016Full, muller2017nonreciprocal, Mascarenhas2016Quantum,fang2017multiple,nefedkin2022dark}. Fratini et al. have shown \cite{Fratini2014FabryPerot} that a system of two atoms, optimally detuned and positioned in a waveguide, can lead to large asymmetries between the signals transmitted along the two directions for a large range of input powers. Importantly, both classical and quantum approaches have the important advantage of not requiring any form of external bias, which makes their practical implementation much easier and compatible, for instance, with conventional platforms for quantum computing and processing. However, so far nonlinearity-induced nonreciprocity in free-space quantum systems has not been investigated to the best of our knowledge.

In this work, we investigate whether it is possible to obtain free-space nonreciprocity in quantum metasurfaces, adding an important tool to this rapidly emerging area of research. We consider a system formed by two parallel 2D arrays each including a large number of atoms, excited by classical plane waves. We demonstrate that, by tailoring the system geometry, it is possible to leverage the quantum nonlinearity of the atoms and obtain large nonreciprocal scattered fields for power levels at which the atomic saturation becomes non-negligible. In particular, we show that the total cross section of the system can be made strongly dependent on the direction of the impinging wave, a feature that cannot be obtained in reciprocal systems. Importantly, here we obtain nonreciprocity by solely engineering the position and detuning of the atoms, without requiring that the electromagnetic environment in which the atoms are placed (i.e., free space) supports nonreciprocal wave propagation \cite{gangaraj2022enhancement}.

\section{System and Model}\label{sec:sys}

\begin{figure}[h!]
  \centering
  \includegraphics[width=0.5 \linewidth]{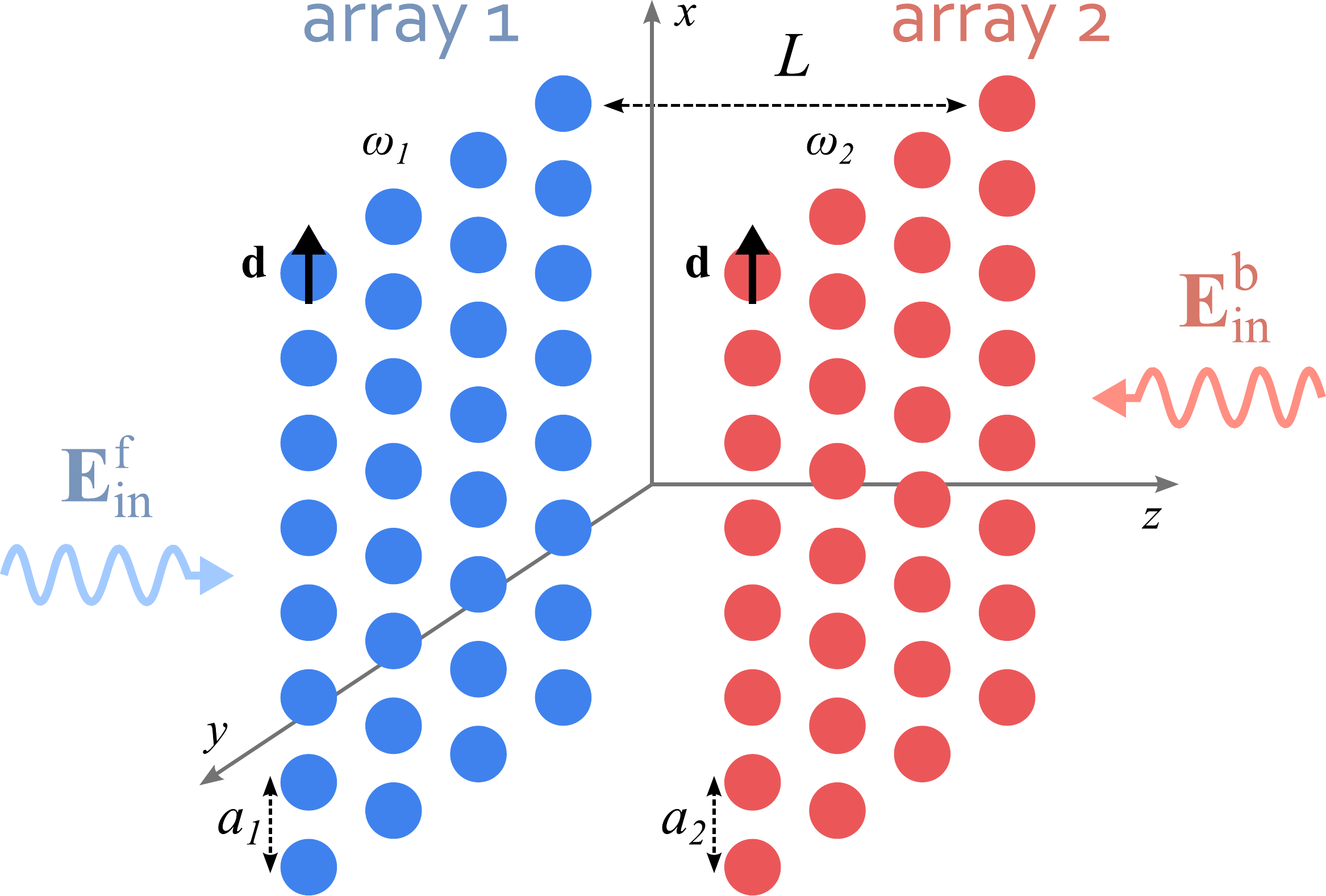}
  \caption{Sketch of two periodic atomic arrays separated by the distance $L$. Both arrays are parallel to the $xy$ plane, and are composed of $N_{\perp} \times N_{\perp}$ identical atoms with frequency $\omega_i$, spatially separated by a lattice constant $a_i$.}
  \label{fig:sys:01}
\end{figure}

We consider a system composed of two squared 2D arrays of two-level atoms (Fig. \ref{fig:sys:01}) The arrays are parallel to each other, orthogonal to the $z$ direction, and separated by a distance $L$.
We assume that each array contains $N_{\perp} \times N_{\perp}$ identical two-level atoms (where $N_{\perp}$ is the number of atoms along each dimension).
In each array ($i=1,2$) the atoms have the same transition frequency $\omega_i$ and they are separated by a lattice pitch $a_i$.
Hence, the total number of atoms in the system is $N\equiv 2\times N_{\perp} \times N_{\perp}$.
For simplicity, we assume that all atoms have the same transition dipole moment $\mathbf{d} = d \mathbf{e}_{x}$, directed along the $x$ axis.
We denote by $|0_j \rangle$ and $|1_j \rangle$ the ground and excited states of the $j$-th atom ($j=1,\ldots, N$).
In the following, we will find it useful to normalize the decay rates of the atomic collective modes by the decay rate of a single atom in free space (with frequency $\omega_i$ and dipole moment magnitude $d$), which is given by  $\gamma^i_{0}  = \frac{\omega_{i}^{3} d^{2}}{3\pi \epsilon_{0} \hbar c^{3}}$ ~\cite{carmichael1999statistical}, where $c$ is the speed of light and $\epsilon_{0}$ is the vacuum permittivity.

We are interested in modeling the response of this system to impinging classical monochromatic plane waves of arbitrary intensities.
For simplicity, we focus here on the case of waves propagating perpendicular to the plane of the arrays, but the formalism can be easily generalized to tilted excitation directions.
The impinging electric field is $\mathbf{E}_{\text{in}}(\mathbf{r})e^{-i\omega_{0}t} = E_{0} \mathbf{e}_{x} \exp(i \mathbf{k}_{0} \mathbf{r})e^{-i\omega_{0}t}$, where the wave vector $\mathbf{k}_{0} = \pm |\mathbf{k}_{0}| \hat{\mathbf{z}}$ has magnitude $|\mathbf{k}_{0}| = \omega_{0}/c = 2 \pi \lambda_{0} /c$ and it points towards either the $+z$ or $-z$ direction. For brevity, we define as \textit{forward} (f) the direction corresponding to propagation along the positive $z$ direction ($\mathbf{k}_{0} = |k_0| \hat{\mathbf{z}}$), and as \textit{backward} (b) the opposite direction ($\mathbf{k}_{0} = -|k_0| \hat{\mathbf{z}}$).

A full quantum description of the system response would require to explicitly describe the temporal dynamics of the infinite free-space modes that the atoms can interact with, which would make any analytical or numerical solution very challenging.
Following standard approaches \cite{bettles_quantum_2020, wild_algorithms_nodate}, it is possible to drastically simplify the system description by applying the Born-Markov approximation.
Assuming that the typical relaxation time of the atoms is much slower than the time required by light to propagate between atoms, the photonic degrees of freedom can be traced out from the full Hamiltonian of the system, resulting into an effective master equation which involves only the atomic degrees of freedom.
In our scenario, the validity of the Born-Markov approximation requires that $1/\gamma_\mathrm{max} \gg d_\mathrm{max}/c$, where $\gamma_\mathrm{max}$ is the largest atomic decay rate and $d_\mathrm{max}$ is the largest inter-atomic distance.
The effective master equation reads~\cite{bettles_quantum_2020, wild_algorithms_nodate}
\begin{equation} \label{eq:sys:01}
	  \begin{split}
    \dot{\hat{\rho}} = \frac{i}{\hbar}\left[ \hat{\rho}, \hat{H}_{S} \right] &+ \sum_{i,j} \frac{\Gamma_{ij}}{2} \left( 2 \hat{\sigma}_{j} \hat{\rho} \hat{\sigma}_{i}^{+} - \hat{\rho} \hat{\sigma}_{i}^{+}\hat{\sigma}_{j} - \hat{\sigma}_{i}^{+} \hat{\sigma}_{j} \hat{\rho} \right)
  \end{split}
\end{equation}
where we have assumed that, in the frequency range of interest, thermal excitations in the environment are negligible, and thus dissipation can only occur \textit{from} the atomic system \textit{into} the environment.
For a general arrangement of atoms in free space excited by a classical EM field, the  effective Hamiltonian (in the frame rotating at the frequency $\omega_{0}$ of the incident field) is ~\cite{asenjo-garcia_exponential_2017, shahmoon_cooperative_2017, bettles_quantum_2020}
\begin{equation} \label{eq:sys:02}
		\hat{H}_{S} = \hbar \sum_{k=1}^{{N}} \left(\Delta_{k} \hat{\sigma}_{k}^{+} \hat{\sigma}_{k} -\Omega_{R}^{k} \hat{\sigma}_{k} - \Omega_{R*}^{k} \hat{\sigma}_{k}^{+} \right) + \hbar \sum_{i\neq j} \Omega_{ij} \hat{\sigma}_{i}^{+} \hat{\sigma}_{j},
\end{equation}
where the indices $k,i,j$ run over all the atoms.
We defined the atom-field detuning, $\Delta_{k} \equiv \omega_{k} - \omega_{0}$, and the interaction constant between the incident field and the $k$-th atom, $\Omega_{R}^{k} \equiv \mathbf{d}\cdot\mathbf{E}_{\mathrm{in}}^{+}(\mathbf{r}_{k})/\hbar$.
For a fixed dipole moment and atom position, $\Omega_{R}^{k}$ is proportional to the impinging field amplitude.
In the system considered in Fig. \ref{fig:sys:01}, due to the planar symmetry of the atomic system and to the plane wave excitation, the coefficients in eq. (\ref{eq:sys:02}) simplify.
In particular, atoms that belong to the same array will have the same detuning (denoted $\Delta^{(1)}$ for the first array and $\Delta^{(2)}$ for the second array), and the same interaction constants, denoted $\Omega_{R}^{(1)}$ and $\Omega_{R}^{(2)}$.
Moreover, $\Omega_{R}^{(1)}$ and $\Omega_{R}^{(2)}$ only differ by a phase factor $e^{ik_0L}$, and we define $|\Omega_{R}|\equiv |\Omega_{R}^{(1)}| = |\Omega_{R}^{(2)}|$.
The terms proportional to $\Gamma_{ij}$ in eq. (\ref{eq:sys:01}) and $\Omega_{ij}$ in eq. (\ref{eq:sys:02}) account for the dipole-dipole dissipative and coherent interactions, respectively, and they read
\begin{align} \label{eq:sys:03}
	\Omega_{ij} =& - \frac{3\pi c}{\omega} \sqrt{\gamma_{0}^{i} \gamma_{0}^{j}} \mathbf{d}_{i}^{*} \mathrm{Re} \left( \overleftrightarrow{\mathbf{G}}(\mathbf{r}_{i}, \mathbf{r}_{j}, \omega)  \right) \mathbf{d}_{j}, \\
  \Gamma_{ij} =& \frac{6\pi c}{\omega} \sqrt{\gamma_{0}^{i} \gamma_{0}^{j}} \mathbf{d}_{i}^{*} \mathrm{Im} \left( \overleftrightarrow{\mathbf{G}}(\mathbf{r}_{i}, \mathbf{r}_{j}, \omega)  \right) \mathbf{d}_{j},
\end{align}
where $\overleftrightarrow{\mathbf{G}}$ is the free-space total Green's tensor,
\begin{equation}\label{eq:sys:03_1}
	\overleftrightarrow{\mathbf{G}}(\mathbf{r}, \mathbf{r}', k) = \left[ \overleftrightarrow{\mathbf{I}} + \frac{1}{k^{2}} \nabla \nabla \right] \frac{e^{i k |\mathbf{r} - \mathbf{r}'|}}{4 \pi |\mathbf{r} - \mathbf{r}'|},
\end{equation}
and $\overleftrightarrow{\mathbf{I}}$ is a unit tensor.

The total electric field, given by the sum of the impinging field and the field scattered by the atoms, can be calculated using the input-output relations~\cite{dung_resonant_2002, asenjo-garcia_exponential_2017},
\begin{equation} \label{eq:res:01}
	\hat{\mathbf{E}}_{\mathrm{tot}}^{+}(\mathbf{r}) = \hat{\mathbf{E}}_{\mathrm{in}}^{+}(\mathbf{r}) + \hat{\mathbf{E}}_{\mathrm{sc}}^{+}(\mathbf{r}).
\end{equation}
The incident field operator is defined as $\hat{\mathbf{E}}^{+}_{\mathrm{in}}(\mathbf{r}) = \mathbf{E}^{+}_{\mathrm{in}}(\mathbf{r}) \hat{I}$, where $\hat{I}$ is an identity operator acting in the Hilbert space of the atoms and $\mathbf{E}^{+}_{\mathrm{in}}(\mathbf{r})$ is the classical field defined above. The scattered field operator is given by
\begin{equation} \label{eq:res:02}
	\hat{\mathbf{E}}_{\mathrm{sc}}^{+}(\mathbf{r}) = \frac{\omega^{2}}{\epsilon_{0} c^{2}} \sum_{j=1}^{N} \overleftrightarrow{\mathbf{G}}(\mathbf{r}, \mathbf{r}_{j}, \omega) \mathbf{d}_{j} \hat{\sigma}_{j}.
\end{equation}
Equations~(\ref{eq:res:01}--\ref{eq:res:02}) allow to calculate the operators associated to the electric field.
Following standard approaches, the expectation value of any operator $\hat{A}$ in a given state of the system, described by the density matrix $\hat{\rho}$, is obtained by $\langle \hat{A} \rangle = \mathrm{Tr}(\hat{A} \hat{\rho})$.
Equations.~(\ref{eq:res:01}--\ref{eq:res:02}) can be used to calculate either the field generated by the ensemble of atoms upon external excitation, or the field generated when the system is prepared in any arbitrary state and there is no external excitation ($\hat{\mathbf{E}}_{\mathrm{in}}^{+}(\mathbf{r})=0$). If the system is prepared in one of its eigenstates, denoted as $ |\psi_{n}\rangle$, the density operator is $\hat{\rho}_{n} = |\psi_{n}\rangle\langle \psi_{n} |$.

\section{Spectral analysis and fields generated by atomic collective states}

In order to understand the scattering properties of this system, it is useful to first study its eigenstates. We initially focus on the single-excitation subspace of the full Hamiltonian.
In order to study also the dissipation in the system, we recast the master equation Eq. (\ref{eq:sys:01}) into a non-Hermitian Hamiltonian,
\begin{equation} \label{eq:sys:04}
	\hat{H}_{\text{eff}} = \hat{H}_{S} - i\hbar \sum_{j, k} \frac{ \Gamma_{jk}}{2} \hat{\sigma}_{j}^{+} \hat{\sigma}_{k}
\end{equation}
where the imaginary part accounts for dissipation.
By introducing the so-called ``quantum jump'' terms, $\hat{\sigma}_{j}^{+} \hat{\sigma}_{k}$, this effective non-Hermitian Hamiltonian becomes formally equivalent to the master equation ~(\ref{eq:sys:01}), as shown in Ref.~\cite{minganti2019quantum}.

\subsection{Symmetric infinite arrays}

For the general case in which the system does not obey parity symmetry  (\textit{i.e.} $\omega_1\neq\omega_2$ and/or $a_1\neq a_2$), it is challenging to diagonalize the Hamiltonian in Eq. (\ref{eq:sys:04}) analytically. In order to gain insights on the system dynamics, we initially assume that the system is symmetric, i.e., $\omega_1=\omega_2 = \omega$, and $a_1= a_2 = a$. Such symmetric double-array system was also investigated in ref.~\cite{guimond_subradiant_2019}.  Assuming infinitely extended lattices, the eigenvalues and eigenstates of the Hamiltonian in eq. (\ref{eq:sys:04}) can be calculated analytically ~\cite{guimond_subradiant_2019}.

Following~\cite{guimond_subradiant_2019}, we label each atom by a vector index $\mathbf{j} = (\mathbf{j}_{\perp}, j_{z})$, where  $\mathbf{j}_{\perp} = [j_{x}, j_{y}]$ are integer numbers identifying the in-plane position of each atom, $\mathbf{R}_{\parallel} (j_{x}, j_{y}) = a(j_x\hat{\mathbf{e}}_x + j_y\hat{\mathbf{e}}_y)$, whereas $j_{z} = 1, 2$ specifies which array each atom belongs to (see Fig.~\ref{fig:sys:01}).
The general single-excitation eigenstate of the Hamiltonian in eq. (\ref{eq:sys:04}), $\hat{H}_{\text{eff}} | \psi_n \rangle =  \mathcal{E}_{n} | \psi_n \rangle$, can be written as
\begin{equation} \label{eq:sys:05}
	| \psi_n \rangle = \sum_{\mathbf{j}} \psi_{n,\mathbf{j}} | e_\mathbf{j} \rangle
\end{equation}
where the sum runs over all the atoms in the system, $ | e_\mathbf{j} \rangle$ is the excited state of the atom identified by $\mathbf{j}$, and $\psi_{n,\mathbf{j}}$ is a complex coefficient denoting the contribution of the $\mathbf{j}$-th atom to the $n$-th eigenstate of the system. Due to parity symmetry and translational invariance, the eigenstates of $\hat{H}_{\text{eff}}$ have a plane-wave nature \citep{asenjo-garcia_exponential_2017,guimond_subradiant_2019}. In particular, each eigenstate can be associated to an in-plane vector $\mathbf{q}_{n}$ defined within the first Brillouin zone of the lattice, and to a parity index $p_n = \pm 1$. Specifically,
\begin{align} \label{eq:sys:06}
	\psi_{n,\mathbf{j}} =& e^{i a \mathbf{j}_{\perp} \cdot \mathbf{q}_{n}} e^{i \pi \frac{p_{n}-1}{2} (j_{z} - 1)} / \sqrt{2 N}
\end{align}
\begin{align} \label{eq:sys:06bis}
  \mathcal{E}_{n} =& - i \frac{3\pi \gamma_{0}}{2(k a)^{2}} \sum_{\mathbf{Q}} \frac{k^{2} - |(\mathbf{q}_{n} - \mathbf{Q}) \cdot \mathbf{d}|^{2}}{k \sqrt{k^{2} - | \mathbf{q}_{n} - \mathbf{Q} |^{2}}} \left( 1 + p_{n} e^{i \sqrt{k^{2} - |\mathbf{q}_{n} - \mathbf{Q}|^{2}} L }\right).
\end{align}
The sums in Eq.~(\ref{eq:sys:06bis}) run over all the vectors $\mathbf{Q} = (2\pi / a) (m_x+m_y)$ of the reciprocal lattice, where $m_{x}$ and $m_{y}$ are integers, and $k \equiv \omega/c$.
The decay rate of each eigenstate can be calculated from the imaginary part of the expression in Eq. \ref{eq:sys:06bis}, $\gamma_{n} \equiv 2 \mathrm{Im}[\mathcal{E}_{n}]$. It is easy to verify that $\gamma_{n}$ is different from zero only when $|\mathbf{q}_{n} - \mathbf{Q} | < k$. In other words, only a finite number of terms in the infinite sum contributes to the decay rate $\gamma_{n}$. The decay rate of the $n$-th eigenstate is therefore given by
\begin{equation} \label{eq:sys:07}
	\gamma_{n} = \frac{3\pi \gamma_{0}}{(k a)^{2}} \sum_{\mathbf{Q}:| \mathbf{q}_{n} - \mathbf{Q} | < k} \frac{k^{2} - |(\mathbf{q}_{n} - \mathbf{Q}) \cdot \mathbf{d}|^{2}}{k \sqrt{k^{2} - | \mathbf{q}_{n} - \mathbf{Q} |^{2}}} \left( 1 + p_{n} \cos \left(\sqrt{k^{2} - |\mathbf{q}_{n} - \mathbf{Q}|^{2}} L \right)\right).
\end{equation}

Equation (\ref{eq:sys:07}) can be further simplified when the lattice constant is smaller than the atomic wavelength, $a < \lambda$, and for eigenstates with zero momentum $\mathbf{q}_{n} = \mathbf{0}$. In this case the two eigenstates, corresponding to the two parity values $p=\pm 1$, have decay rates  $\gamma_{\pm 1} = \frac{3\pi \gamma_{0}}{(k a)^{2}} \left( 1 \pm \cos(k L) \right)$.
In the specific case where $k L  = m \pi$ with $m$ an integer, we obtain  $\gamma_{-1} = 0$ and $\gamma_{+1} = \frac{6\pi \gamma_{0}}{(k a)^{2}} = \frac{3}{2\pi}\left(\frac{\lambda}{a}\right)^2\gamma_0$. Here, the system has a \textit{dark} eigenstate, characterized by zero decay rate, and a \textit{bright} eigenstate, with a decay rate which is finite and larger than $\gamma_0$ (because of the  $a < \lambda$ assumption). In the following, we will use the subscripts D and B to denote quantities pertaining to the dark and bright states, respectively. The dark and bright states are the result of collective effects in the atomic array, leading to constructive and destructive interference of the collective emission.
As we will discuss later, the existence of these states is the key property paving the way to nonreciprocal effects in the scattered fields.

\subsection{Symmetric finite arrays}

In the previous section we have considered quantum metasurfaces formed by infinitely extended periodic arrays, which allowed us to calculate analytically the eigenstates of the atomic ensemble and to show the existence of dark and bright states. In the remainder of the paper we will focus on systems of finite size. This is due to two reasons: first, we want to be sure that the predicted effects are observable in experimentally feasible atomic systems, which are currently limited to squared arrays with $N_{\perp} \times N_{\perp} \leq 14\times 14$ \cite{rui_subradiant_2020}. Second, considering finite and small arrays will facilitate the analysis of their nonlinear dynamics. When multiple collective excitations are created in the atomic system, analytical results are very challenging to obtain and we will therefore resort to numerical calculations, which are only feasible for small systems.

\begin{figure}[t!]
  \centering
  \includegraphics[width=0.5 \linewidth]{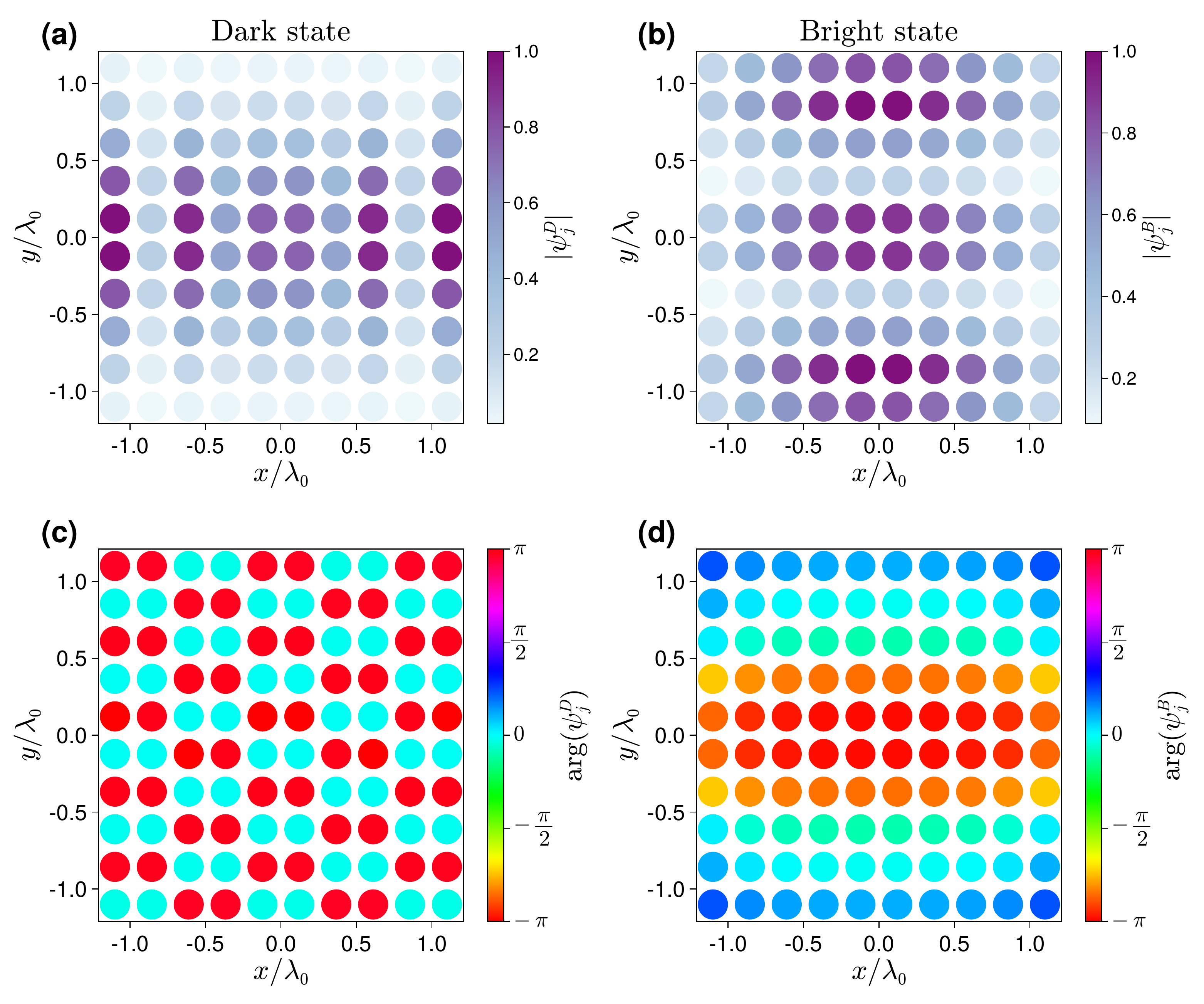}
  \caption{Dark (left panels) and bright (right panel) eigenstates of a two-array system with $N_{\perp} \times N_{\perp} = 10\times 10$ atoms in each array obeying mirror symmetry ($a_1=a_2=\lambda_{0}/4,\omega_1=\omega_2$) for $L=0.6 \lambda_{0}$. Panel (a) and (c) show the amplitude $|\psi_{j}|$ and phase $\mathrm{arg}(\psi_{j})$ of the dark eigenstate at each atom in the first array. Panels (b) and (d) show the same quantities for the bright state. The amplitudes are the same in the second array, while all phases are shifted by $\pi$ due to the parity parameter $p_n$ in eq. (\ref{eq:sys:06}) ($p_{B,D} = -1$ in this case).}
  \label{fig:res:01}
  \centering
  \includegraphics[width=0.5 \linewidth]{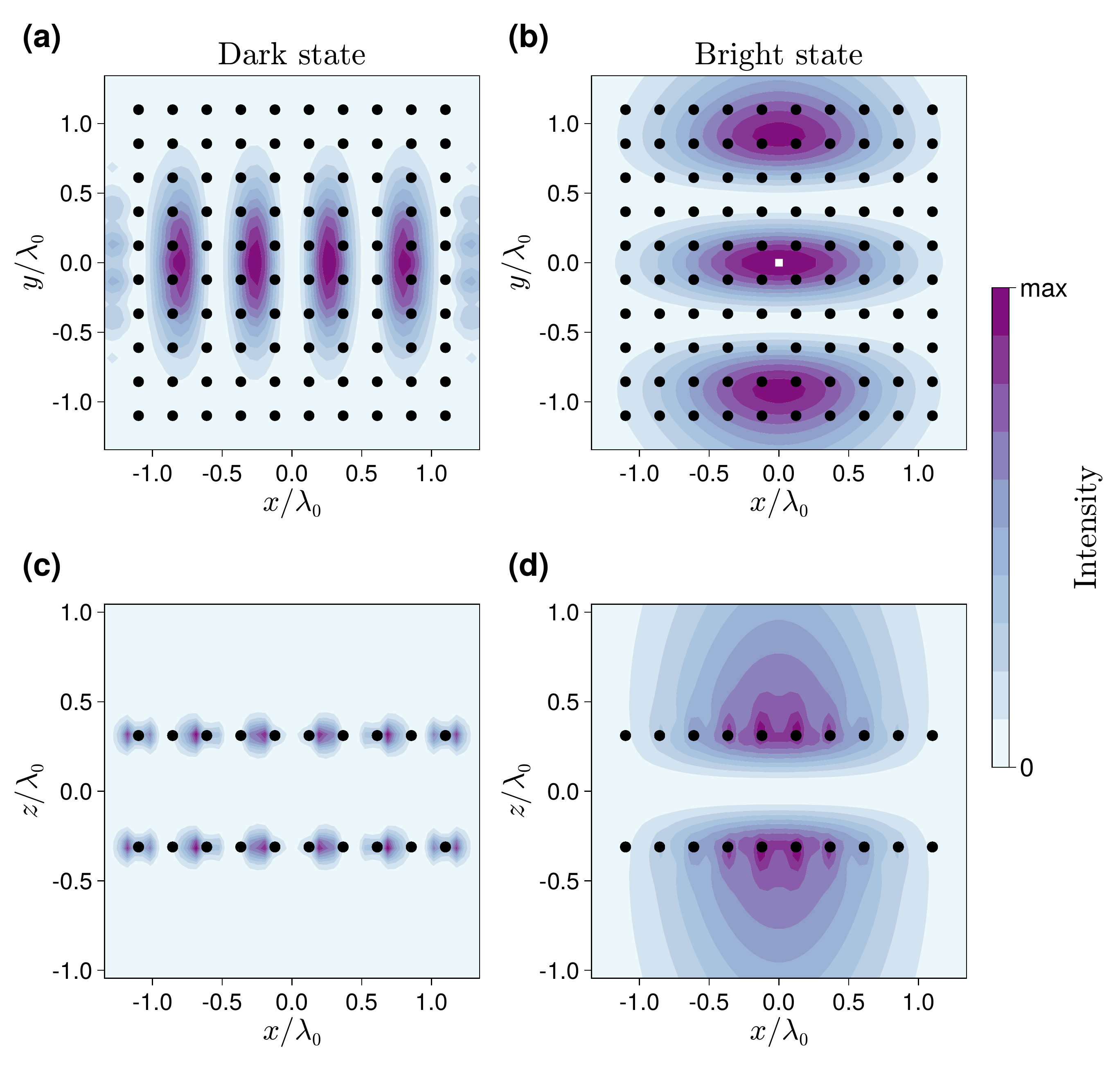}
  \caption{Field intensity in the dark and bright state: (a), (b) slices in $xy$ plane; (c), (d) slices in $xz$ plane.}
  \label{fig:res:02}
\end{figure}

As discussed in ~\cite{guimond_subradiant_2019}, in finite symmetric arrays the collective decay rates are generally different from the ones calculated in infinite arrays eq. (\ref{eq:sys:07}). One can still identify \textit{bright} and \textit{dark} states by looking at eigenstates whose decay rate are much larger or smaller, respectively, than the decay rate of a single atom in free space. However, the decay rate of dark states in finite systems remains always larger than zero. In Ref.~\cite{guimond_subradiant_2019}, the authors addressed this issue and minimized the decay rate of a dark state by assuming that the arrays are not planar but instead possess a Gaussian-like curvature, calculated based on the phase profile of a Gaussian mode.
This allowed them to find analytical expressions for decay rates and frequency shifts of the eigenstates, and to express the eigenmodes as Hermit-Gaussian modes with $|\mathbf{q}_{n}|$ close to $0$.

In order to keep our system more realistic, we restrict ourselves to planar arrays, considering collective modes with arbitrarily large $|\mathbf{q}_{n}|$, and we calculate numerically the single-excitation spectrum of the effective Hamiltonian in eq. (\ref{eq:sys:04}). As an example, in Fig.~\ref{fig:res:01} we consider a system of two identical arrays, each of them with $N_{\perp} \times N_{\perp} = 10\times 10$ atoms and we plot the amplitudes of the dark state (defined here as the eigenstate with the smallest decay rate) and the bright state (defined here as the eigenstate with the largest decay rate) at each atomic location in the first array.

Assuming that the system is prepared in one of these eigenstates (and in absence of any impinging field), we can calculate the electric field intensity generated by the quantum metasurfaces. By using Eq.~(\ref{eq:res:02}), the field intensity generated at any point is $I(\mathbf{r}) \propto \sum_{j,j'}\mathbf{d}_{j}^{*} \left(\overleftrightarrow{\mathbf{G}}^{*}(\mathbf{r}-\mathbf{r}_{j}, \omega) \overleftrightarrow{\mathbf{G}}(\mathbf{r}-\mathbf{r}_{j'}, \omega) \right) \mathbf{d}_{j'} \langle \hat{\sigma}^{+}_{j} \hat{\sigma}_{j'} \rangle$.
The field distribution in the $xy$ plane is expected to follow the pattern shown in Fig.~\ref{fig:res:01}, whereas the distribution in the $xz$ plane gives us new information about the character of emission of the system in the dark and bright states.

Figure~\ref{fig:res:02} shows the field intensity generated by the bright and dark states in the plane of the array ($xy$) and in a plane perpendicular to it ($xz$). The in-plane field profiles (Figs.~\ref{fig:res:02}a-b) bears similarity to a Hermit-Gaussian mode~\cite{guimond_subradiant_2019}. As expected, the dark mode (i.e., an eigenstate with almost-zero decay rate) is associated to an almost-zero field far from the arrays (Fig.~\ref{fig:res:02}c). On the other hand, the bright mode generates a strong field along the two directions perpendicular to the arrays (Fig.~\ref{fig:res:02}d).

\begin{figure}[h!]
  \centering
  \includegraphics[width=0.5 \linewidth]{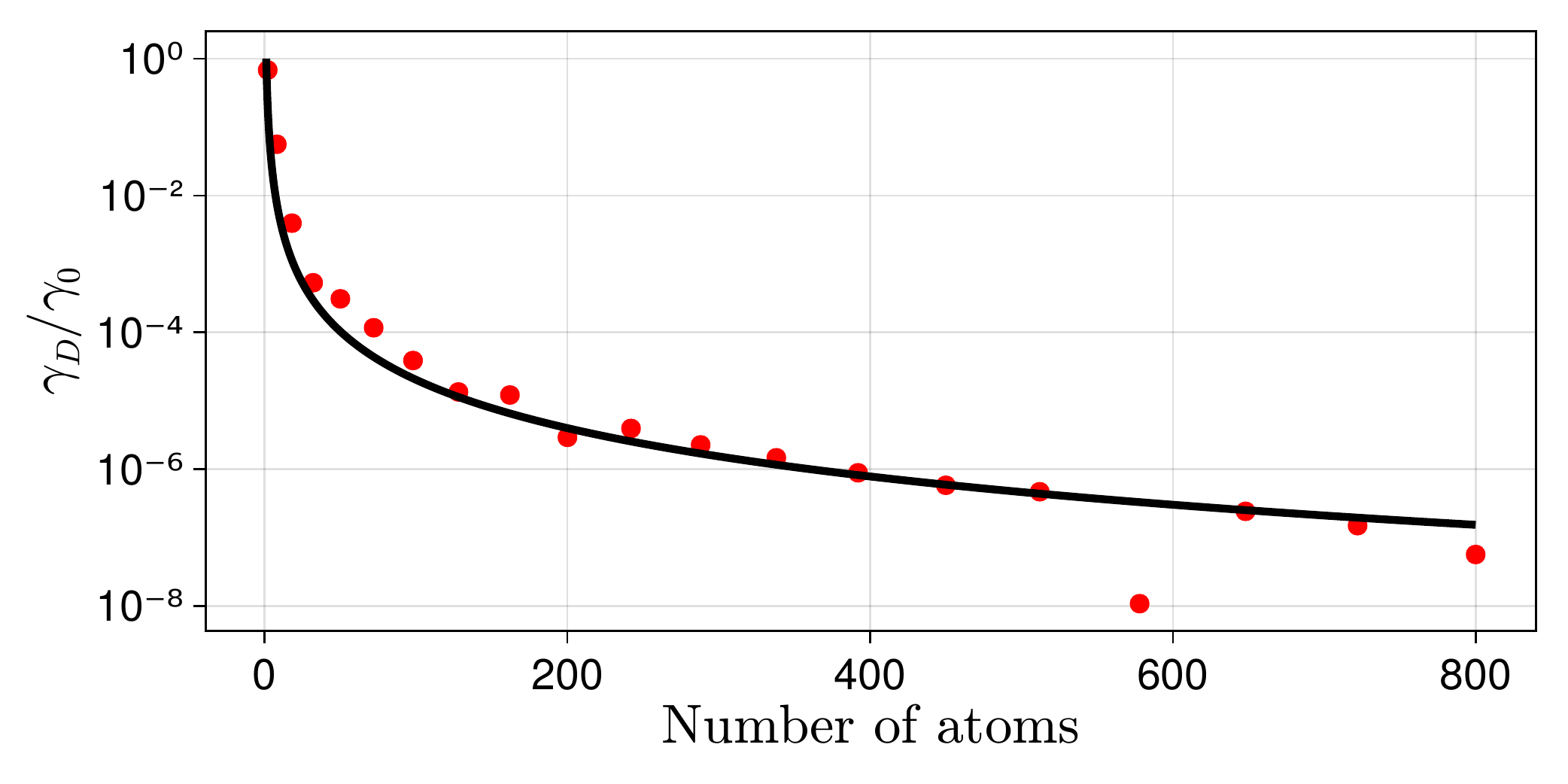}
  \caption{Dependence of the decay rate of the dark state on the number of atoms in the system. The black line is the fit $\gamma_{D} \sim N^{-\alpha}$, $\alpha = 2.5$.}
  \label{fig:res:03}
\end{figure}

While the decay rate of dark states in a finite atomic system will always be non-zero, it can be made arbitrarily small by increasing the number of atoms. In particular, as discussed in \cite{asenjo-garcia_exponential_2017} the decay rate of the dark state is expected to scale as $\gamma_{D} \sim N^{-3}$ for collective modes at the edge of the Brillouin zone, where $N$ is the total number of atoms in the system.
In Fig.~\ref{fig:res:03} we show the dependence of $\gamma_{D}$ on $N$, while all other parameters are left constant (see caption to Fig.~\ref{fig:res:01}).
We note that the decay rates of bright states fulfill the condition of applicability of the Born-Markov approximation discussed above, i.e., $1 / \gamma_{D} \gg d_{\mathrm{max}} / c$, for all the values of $N$ considered here.

For the largest arrays considered here ($N_{\perp} \times N_{\perp} = 20\times 20$, $N=800$), $\gamma_{D}$ is about $10^{-8}$ smaller than the decay rate of a single atom in free space ($\gamma_{0}$).
The black curve in Fig.~\ref{fig:res:03} shows the fit of the decay rates of the dark state, $\gamma_{D} \sim N^{-\alpha}$, with $\alpha \approx 2.5$. This value is smaller than the one found in the case of one array \cite{asenjo-garcia_exponential_2017}, which we tentatively attribute to the interactions between the arrays at the parameters of choice. In a different geometry, $L = \lambda_{0}$ and $a = \lambda_{0} / 5$, we found that $\alpha = 3$.
As suggested in Refs.~\cite{asenjo-garcia_exponential_2017, guimond_subradiant_2019}, such extremely subradiant atomic systems can be used as a quantum memory in network and information processing applications, and protocols for writing and reading states in these systems have been proposed~\cite{guimond_subradiant_2019,reitz_cooperative_2022}.

\section{Nonreciprocal total cross section in asymmetric systems}\label{sec:results}
\subsection{Arrays with few atoms -- Quantum analysis} \label{subsec:quantum}

After having elucidated the spectral properties of symmetric systems, we now investigate under what conditions a two-array system can lead to nonreciprocity. As discussed in the introduction, we expect, based on general considerations \cite{cotrufo2021nonlinearityPart1}, that a system can become electromagnetically nonreciprocal when geometric asymmetry is combined with a nonlinear optical response. When restricted to the single-excitation subspace, the dynamics of the two-array system is fully linear, and it is analogous to a system of classical dipoles. The two-level atoms, however, feature a saturable absorption. A nonlinear response is therefore obtained at sufficiently high excitation powers. The required geometric asymmetry can be obtained by relaxing the assumption that $a_{1}=a_{2}$ and/or $\omega_{1}=\omega_{2}$. Note that, when considering impinging fields with arbitrarily large power, we cannot use low-power approximations~\cite{asenjo-garcia_exponential_2017,shahmoon_cooperative_2017} which neglect nonlinear terms.

In order to quantify the degree of nonreciprocity for a given geometry and input power we calculate the total cross section of the  system (for both excitation directions), which can be readily obtained via the optical theorem. Following the formalism described above, we assume that the system is excited by a plane wave with spatial field distribution $\mathbf{E}_{\text{in}}(\mathbf{r}) = E_{0} \mathbf{e}_{0} \exp(i \mathbf{k}_{0} \mathbf{r})$, where the impinging propagation direction is dictated by the wave vector $\mathbf{k}_{0} = \pm|k_0| \hat{\mathbf{z}}$. At large distances ($r \gg \lambda_{0}$), the field scattered along the propagation direction, $\mathbf{E}_{\text{sc}}(\mathbf{r}=\pm r\hat{\mathbf{z}})$, can be written as an asymptotic spherical wave,
\begin{equation} \label{eq:res:03bis}
\mathbf{E}_{\text{sc}}(\mathbf{r}=\pm r\hat{\mathbf{z}}) \rightarrow \frac{e^{ik_{0}r}}{r} E_{0} \mathbf{f}(\mathbf{k}=\mathbf{k}_{0}),
\end{equation}
where $\mathbf{f}(\mathbf{k})$ is the \textit{scattering amplitude} along the direction defined by $\mathbf{k}$.
By applying the optical theorem~\cite{jackson1999classical}, the total extinction cross section of the system can be then calculated as
\begin{equation} \label{eq:res:03}
	\sigma_{\mathrm{tot}}(\mathbf{k}_{0}) = \frac{4\pi}{k_{0}} \mathrm{Im}(\mathbf{e}_{0}^{*}\cdot \mathbf{f}(\mathbf{k} = \mathbf{k}_{0})).
\end{equation}
Note that in the following text we use total cross section implying the total extinction cross section. In our model, the scattering amplitude can be found by solving numerically the master equation (\ref{eq:sys:01}) and calculating the expectation values of the dipole operators in steady-state. After that, the expectation value of the scattered field $\mathbf{E}_{\mathrm{sc}}^{+}$ at a point very far from the quantum metasurfaces is found using Eq.~(\ref{eq:res:02}), and then the scattering amplitude is extracted using eq. (\ref{eq:res:03bis}).

We are interested in quantifying the nonreciprocal extinction of the system, i.e., how differently the system scatters and absorbs waves propagating along opposite directions. Following the forward (f) / backward (b) notation defined above, we define the forward total cross section,  $\sigma_{\mathrm{tot}}^{f} \equiv \sigma_{\mathrm{tot}}(|k_0| \hat{\mathbf{z}})$ and the backward total cross section,  $\sigma_{\mathrm{tot}}^{b} \equiv \sigma_{\mathrm{tot}}(-|k_0| \hat{\mathbf{z}})$. This allows us to the quantify the degree of nonreciprocity by the \textit{nonreciprocal extinction efficiency}, defined as
\begin{equation} \label{eq:res:04}
  \mathcal{M} \equiv \max(\sigma_{\mathrm{tot}}^{f}, \sigma_{\mathrm{tot}}^{b}) \left|\frac{\sigma_{\mathrm{tot}}^{f}-\sigma_{\mathrm{tot}}^{b}}{\sigma_{\mathrm{tot}}^{f}+\sigma_{\mathrm{tot}}^{b}}\right|.
\end{equation}
For a reciprocal system, $\mathcal{M} = 0$ always.
In order to find an optimal geometry, we employed an optimization algorithm based on gradient descent to find the optimal system parameters which maximize the function $\mathcal{M}^2$ (we used the square of $\mathcal{M}$ in order to get a smooth objective function and its derivative). In the optimization, we fix $N_{\perp}$ to a given value, and we optimize the following parameters: the lattice constant of the array 1, $a_{1}$, the difference between the array lattice constants, $\delta \equiv a_{2} - a_{1}$, the distance between arrays $L$, the frequency detuning relative to the frequency of the incident field $\Delta = 2 (\omega_{0} - \omega_{1}) = 2(\omega_{2} - \omega_{0}) = \omega_{2} - \omega_{1}$ (where we assume that the incident field frequency always lies in the middle between $\omega_{1}$ and $\omega_{2}$) and the amplitude of the incident field $E_{0}$. For numerical convenience, in the following plots we normalize the total cross sections and the function $\mathcal{M}$ by the total cross section of a single atom with transition frequency $\omega_{0}$, spontaneous emission rate $\gamma_{0}$ and dipole moment $\mathbf{d}$, placed in free space, which is denoted $\sigma^0_{\mathrm{tot}}$ (see Appendix A for the derivation).

\begin{figure}[h!]
  \centering
  \includegraphics[width=0.8 \linewidth]{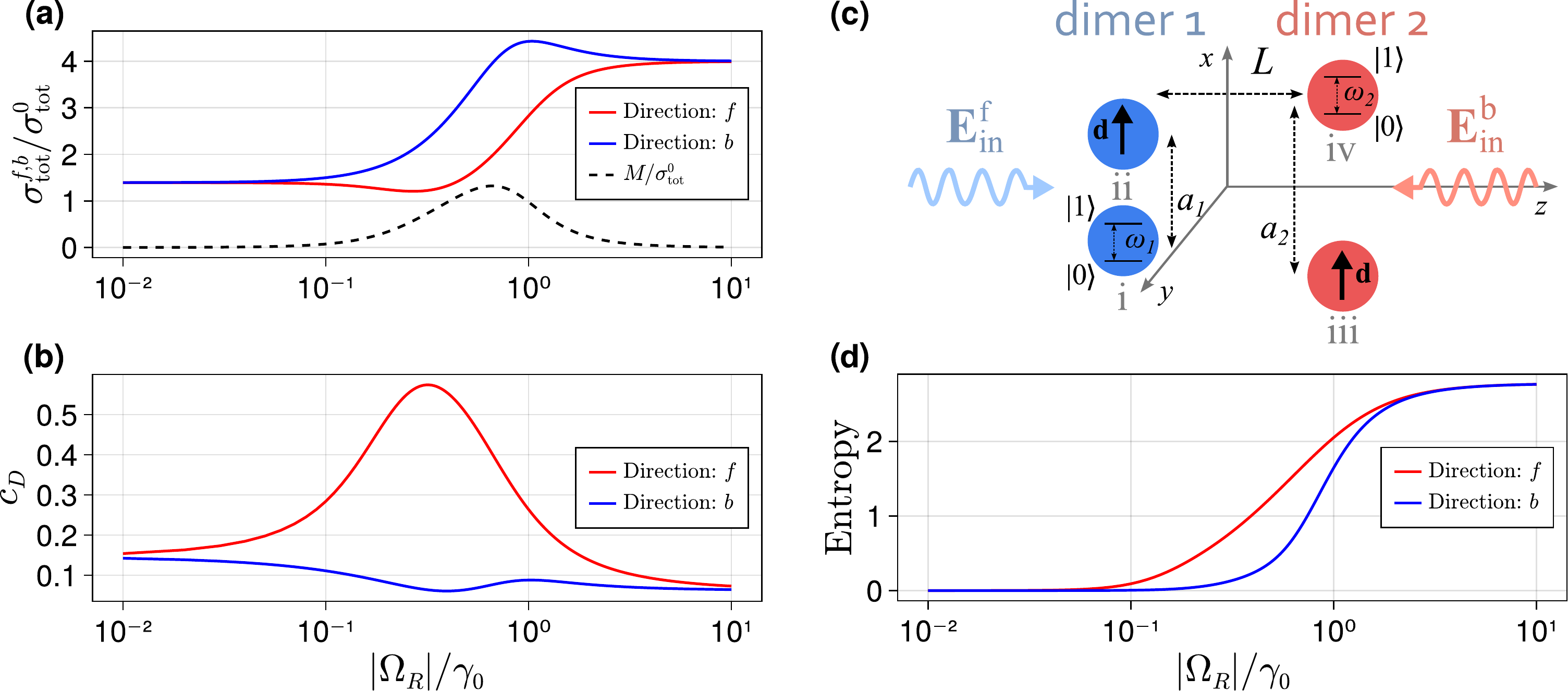}
  \caption{(a) Total cross section, (b) population of the dark state and (d) von Neumann entropy of the two-dimer system (with total number of atoms $N = 4$) when exciting along the forward direction (red line) and the backward direction (blue line), versus the amplitude of the incident field. The two-dimer system is sketched in panel (c). The black dashed line in panel (a) shows the corresponding value of the  nonreciprocal extinction efficiency $\mathcal{M}$. The system parameters are $\Delta = -\gamma_{0}$, $\delta = a_{2} - a_{1} = \lambda_{0} / 10$, $a_{1} = \lambda_{0}/3$, $L = \lambda_{0} / 10$.}
  \label{fig:res:06}
\end{figure}

\begin{figure}[h!]
  \centering
  \includegraphics[width=0.95 \linewidth]{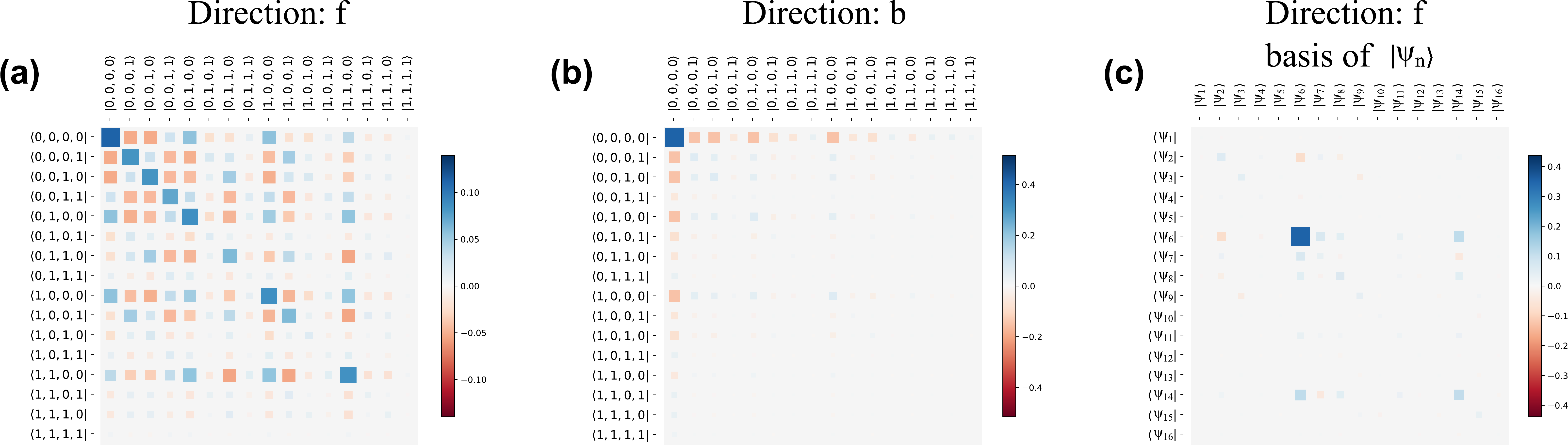}
  \caption{The elements of the steady-state density matrices when the two-dimer system is excited along the forward (a) and backward (b) direction. In panels (a) and (b) the density matrix is plotted in the basis of the uncoupled atomic states (as described in the text). In panel (c) we plot the same density matrix as in panel (a), but with respect to the basis of the system eigenstates, $| \psi_{n} \rangle$, enumerated from $1$ to $16$. The dark state corresponds to the eigenstate $|\psi_5 \rangle$. The system parameters are $\Delta = -\gamma_{0}$, $\delta = a_{2} - a_{1} = \lambda_{0} / 10$, $a_{1} = \lambda_{0}/3$, $L = \lambda_{0} / 10$, $|\Omega_R| = \gamma_0 / 2$.}
  \label{fig:res:05}
\end{figure}

We start by optimizing the geometry and input power to achieve large nonreciprocal extinction in a simple system composed of two dimers. That is, each array is composed of only two atoms separated along the $x$ axis by a distance $a_i$, and the two dimers are separated by a distance $L$ along the $z$ axis, see Fig.~\ref{fig:res:06}(c).
We found that a global maximum of $\mathcal{M}^{2}$ occurs for $\Delta = -\gamma_{0}$, $\delta = a_{2} - a_{1} = \lambda_{0} / 10$, $a_{1} = \lambda_{0}/3$, $L = \lambda_{0} / 10$, $|\Omega_{R}| = \gamma_{0}/ 2$. Fig.~\ref{fig:res:06}(a) shows, for a system with these optimized parameters, the total cross sections for both excitation directions versus the impinging power. A clear nonreciprocal region, whereby the cross sections $\sigma_{\mathrm{tot}}^{f}$ and $\sigma_{\mathrm{tot}}^{b}$ are markedly different, is visible. As expected, at low impinging powers the total cross sections $\sigma_{\mathrm{tot}}^{f}$ and $\sigma_{\mathrm{tot}}^{b}$ are the same. This is due to the fact that, for low powers, the population inversion of the atoms is negligible ($\langle \hat{\sigma}_{j}^{z} \rangle \approx -1$). In this scenario the atoms behave as a system of classical linear dipoles, which is bound to be reciprocal. Moreover, the total cross sections are identical also at high powers. In this case the atoms of both arrays are completely saturated ($\langle \hat{\sigma}_{j}^{z} \rangle \approx 0$), and the whole system becomes fully transmissive along both directions. The black dashed curve in Fig.~\ref{fig:res:06}(a) shows that the maximum of the function $\mathcal{M}$ is obtained for impinging powers such that $|\Omega_{R}| \approx \gamma_{0}/2$.

To gain more insight on the system dynamics, we look at the steady-state density matrices of the dimer system for the two opposite propagation directions, denoted $\hat{\rho}_{f}$ and $\hat{\rho}_{b}$, and for a value of impinging powers corresponding to the largest nonreciprocal extinction. The elements of these matrices are shown in Fig.~\ref{fig:res:05}(a,b), with respect to the basis of the uncoupled atomic states. Specifically, the vector $|s_{\mathrm{i}}, s_{\mathrm{ii}}, s_{\mathrm{iii}}, s_{\mathrm{iv}}\rangle$ denotes a state where each of the 4 atoms of the two-dimer system is either in the ground ($s_j = 0$) or the excited ($s_j = 1$) state. Figure ~\ref{fig:res:05}a shows that, for forward excitation, the density matrix has a non-trivial structure and it is a superposition of several states of the uncoupled atomic basis.
The density matrix for the backward excitation direction (Figure ~\ref{fig:res:05}b), and for the \textit{same} impinging power, has, in contrast, a simple structure in which the ground state $|0,0,0,0\rangle$ is the most populated one. This gives a first hint to explain the nonreciprocal total cross sections observed in Fig. \ref{fig:res:06}(a). Similar to situations involving few atoms in waveguides~\cite{chang2007single, muller2017nonreciprocal, nefedkin2022dark}, a larger ground state population leads to stronger reflection when the atomic system is excited on resonance. A larger reflection implies a larger total cross section, thus explaining the asymmetry observed in Fig. \ref{fig:res:06}(a).

Further insights can be gained by plotting the density matrix for forward excitation, $\hat{\rho}_{f}$, with respect to the basis of the eigenstates of the system, denoted $|\psi_j\rangle$ ($j=1, \ldots, 16$), which are obtained by diagonalizing the corresponding non-Hermitian Hamiltonian. The result, displayed in Fig.~\ref{fig:res:05}(c), clearly shows that for forward excitation the system populates almost exclusively the eigenstate $|\psi_{5} \rangle$. By calculating the corresponding eigenvalues of (\ref{eq:sys:04}) (not shown here), we verified that $|\psi_{5} \rangle$ is a dark state, with decay rate $\gamma_{D} = 0.6 \gamma_{0}$.
Thus, nonreciprocity is intimately linked to an asymmetric state population. In a certain power range, the steady state of the system evolves towards a dark state when excited along the forward direction, while it evolves into the ground state when excited from the opposite direction.
This mechanism is analogous to what described in Refs.~\cite{muller2017nonreciprocal, hamann2018nonreciprocity, nefedkin2022dark} for the case of two atoms in a waveguide, whereby different population transfer paths are realized due to interference.

Such asymmetric state population is more clearly displayed in Fig.~\ref{fig:res:06}(b), which shows the population of the dark state, $c_{D}$, versus impinging power for the forward and backward directions of excitation. For forward excitation direction the population of the dark state reaches $0.6$ within the range of powers where nonreciprocity is large, while it drops to low values at both low and high impinging powers. In the same range of powers, the dark state population is below $0.1$ for backward excitation, and it remains low throughout the range of $E_{0}$.

The dynamics of the system can be further analyzed with the help of the von Neumann entropy which, for any arbitrary state $\hat{\rho}$, is defined as
\begin{equation} \label{eq:res:05}
	S_{\mathrm{vN}} = - \mathrm{Tr}(\hat{\rho} \ln (\hat{\rho})).
\end{equation}
The entropy $S_{\mathrm{vN}}$ quantifies the degree of mixing of the state $\hat{\rho}$. In particular, it is equal to zero when the system is in a pure state, and it increases as the state $\hat{\rho}$ becomes a more and more complicated mixture of pure states. In our system the maximum value of entropy is $S_{\mathrm{vN}}= \ln(2^{N})$, where $N$ is the total number of atoms, corresponding to a maximally mixed state. The von Neumann entropy versus impinging power for both excitation directions is shown in Fig.~\ref{fig:res:06}(d). For low powers, $S_{\mathrm{vN}} \approx 0$ for both directions, corresponding to the fact that system remains in the (pure) ground state.
As the impinging power increases, the entropy increases as well, but in a different way for the two impinging directions. For forward excitation (red line in Fig.~\ref{fig:res:06}(d)) the dark state gets partially populated at intermediate powers ($c_{D} = 0.6$). However, the steady-state of the system in this scenario is not pure, since additional states are populated. This increases the entropy. When exciting along the backward direction (blue line in Fig.~\ref{fig:res:06}(d)), at the same intermediate power level, the system is instead almost entirely localized in the ground state, which results in a lower entropy. At large impinging powers, the entropy tends to its maximum value, $S_{\mathrm{vN}}= \ln(2^{N})\approx 2.77$ for both directions of excitation, highlighting that the system is in a maximally mixed state, see also Fig.~\ref{fig:app:02} in Appendix B.

\begin{figure}[h!]
  \centering
  \begin{subfigure}{0.8\textwidth}
    \includegraphics[width=0.9\linewidth]{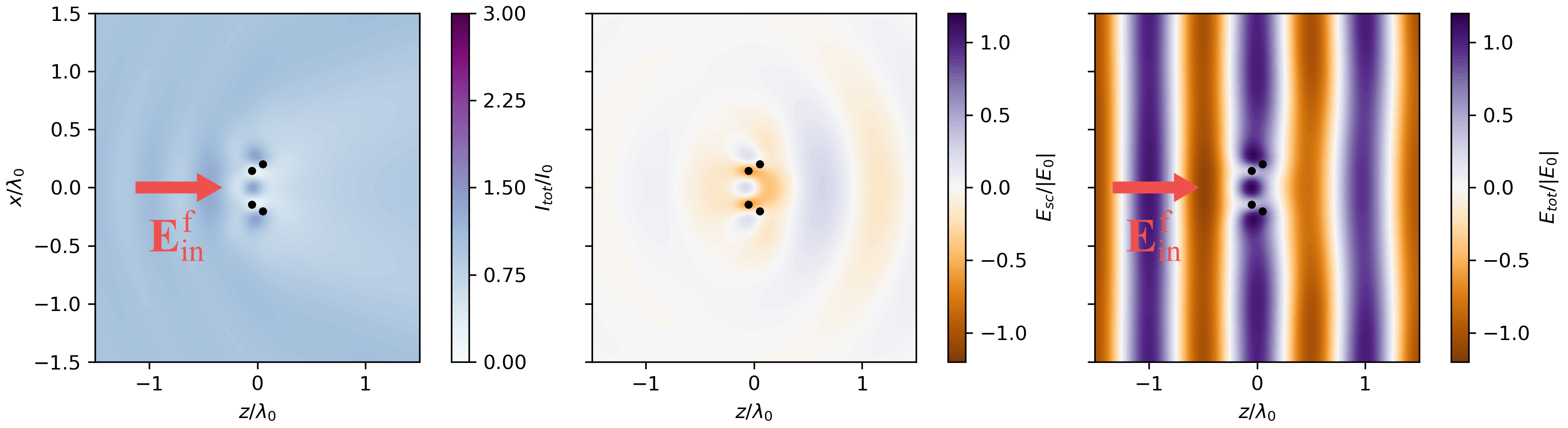}
    \caption{Forward direction.}
  \end{subfigure}
  \vfill
  \begin{subfigure}{0.8\textwidth}
    \includegraphics[width=0.9\linewidth]{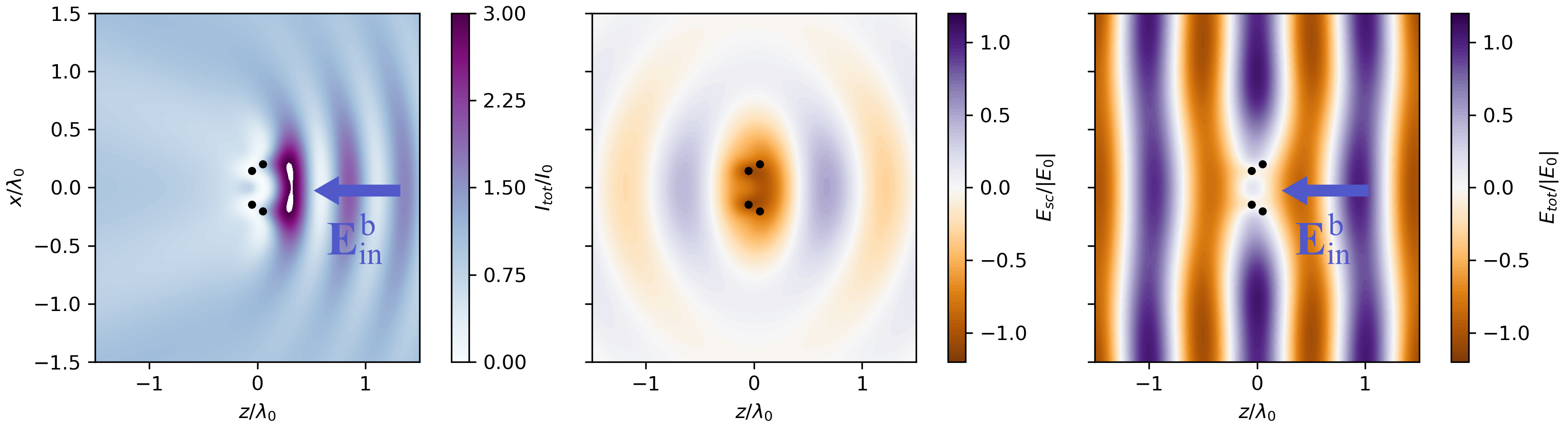}
    \caption{Backward direction.}
  \end{subfigure}
  \caption{Total field intensity, real part of $E_\text{sc}^{x}$ and real part of $E_\text{tot}^{x}$ of 2-array system, with each array containing $N_{\perp}\times N_{\perp} = 2\times2$ atoms, for forward (panel a) and backward (panel b) excitation direction Parameters: $\Delta = -\gamma_{0}$, $\delta = a_{2} - a_{1} = \lambda_{0} / 10$, $a_{1} = \lambda_{0}/3$, $L = \lambda_{0} / 10$, $|\Omega_{R}| = \gamma_{0}/ 2$.}
  \label{fig:res:04}
\end{figure}

The system considered in Figs \ref{fig:res:06}-\ref{fig:res:05} was obtained by optimizing the geometry and parameters of the atomic ensemble for a given input frequency $\omega_0$. As discussed above, here nonreciprocity is due to the asymmetrical excitation of a dark state. Thus, the nonreciprocal behaviour is expected to depend strongly on the excitation frequency (for a given set of system parameters). This is confirmed by the calculations shown in Appendix B, where we calculate the nonreciprocal efficiency $\mathcal{M}$ for different values of the input frequency $\omega_0$.

Having verified the emergence of nonreciprocal extinction in a small system composed of two dimers, we now consider a slightly bigger system, composed of two squared arrays with $N_{\perp}\times N_{\perp} = 2\times2$ atoms each. For this system we are unable to find the global maximum of the $\mathcal{M}^{2}$ function in the entire parameter space, due to the higher numerical complexity. However, we found that the same optimal parameters which optimize the two-dimer system (see the caption of Fig.~\ref{fig:res:06}) lead to large nonreciprocal effects also in this larger system. In Fig.~\ref{fig:res:04} we show the total field intensity, the real part of scattered field ($x$-component) and the real part of total field ($x$-component) when this larger system is exited along the forward or backward direction, for the power level corresponding to $|\Omega_{R}| = \gamma_{0} / 2$ (a similar plot for the two-dimer system can be found in the Appendix B, Fig.~\ref{fig:app:01}). A clear nonreciprocal behaviour can be seen in Fig.~\ref{fig:res:04}: for a wave incident along the forward direction the magnitude of the scattered field is much smaller than the case in which the same wave propagates along the backward direction.

We emphasize that, in all the systems considered in this work, the difference between the total cross sections (i.e. the fact that $\sigma_{\mathrm{tot}}^{f}>\sigma_{\mathrm{tot}}^{b}$ within a certain power range) is solely dictated by the geometry of the arrays and by the distance between them, which in turns introduces a phase shift in the transmitted field $\exp(i k_{0} L)$. In particular, the direction along which the total cross section is the largest can be switched by simply changing the value of $L$. For example, for the two-dimer system the total cross section along the backward excitation becomes larger than the one along the forward direction when, for example, $L > \lambda_{0}/2$.

In conclusion, nonreciprocity arises in the system due to the realization of different routes of population transfer depending on the excitation direction. When the system is excited along the forward direction the population is partially trapped in the dark state, whereas for the opposite direction the system remains in the ground state.
This asymmetric population in the system's steady state (Fig.~\ref{fig:res:05}) leads to nonreciprocal extinction, which manifests itself in different total cross sections, Fig.~\ref{fig:res:06}.

\subsection{Large arrays -- Semiclassical approach}\label{subsec:semiclass}

\begin{figure}[h!]
  \centering
  \begin{subfigure}{0.8\textwidth}
    \includegraphics[width=0.9\linewidth]{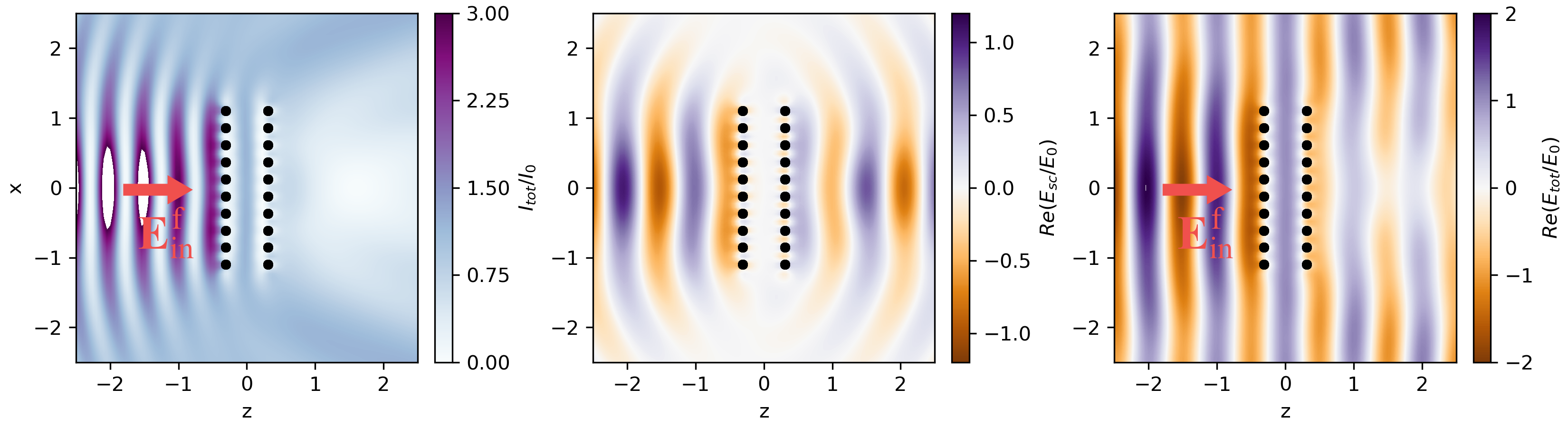}
    \caption{Forward direction.}
  \end{subfigure}
  \vfill
  \begin{subfigure}{0.8\textwidth}
    \includegraphics[width=0.9\linewidth]{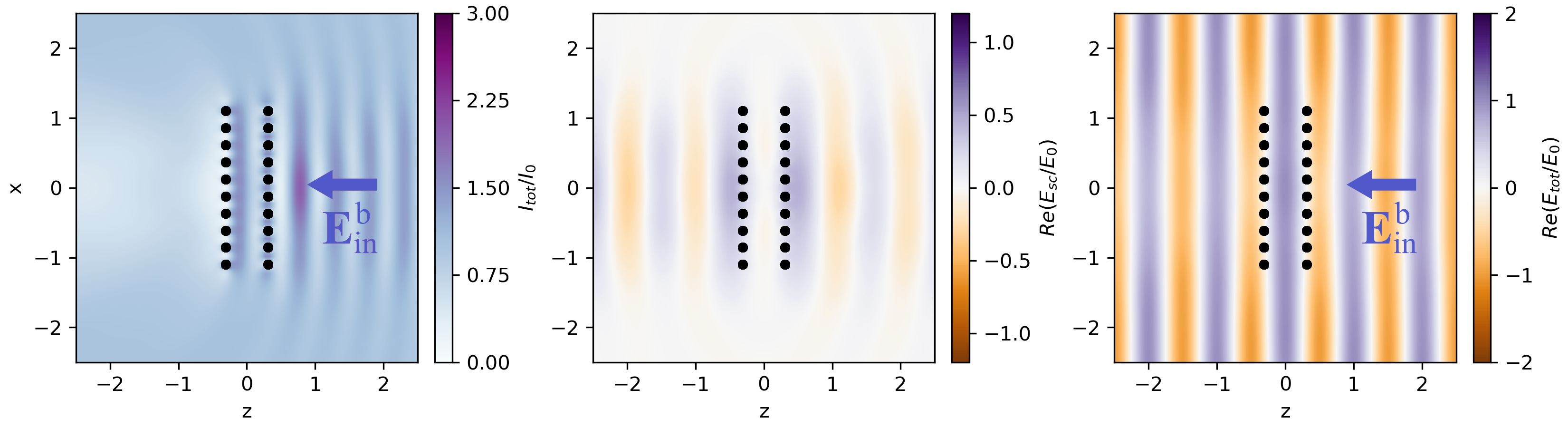}
    \caption{Backward direction.}
  \end{subfigure}
  \caption{Total field intensity, real part of $E_\text{sc}^{x}$ and real part of $E_\text{tot}^{x}$ upon forward (panel a) and backward (panel b) excitation of a 2-array system, each of them composed by $N_{\perp} \times N_{\perp} = 10 \times10 $ atoms. Parameters: $\delta = a_{2} - a_{1} = 0$, $a_{1} = \lambda_{0} / 4$, $\Delta = 2 \gamma_{0}$, $L = 0.6 \lambda_{0}$, $|\Omega_{R}| = 2 \gamma_{0}$.}
  \label{fig:res:07}
\end{figure}

In the previous section we have focused on systems composed of few atoms. Keeping the number of atoms small ($ N \leq 10 $) allowed us to numerically solve the full master equation~(\ref{eq:sys:01}) without any approximation. However, as the number of atoms increases, the exponential increase of the dimension of the Hilbert space prevents us from using the full quantum formalism to calculate the dynamics of larger arrays. This is particularly challenging for the scenario considered in this work: as we are considering large excitation powers, we cannot truncate the Hilbert space to the single-excitation subspace. A common way to address this challenge, and to be able to describe systems composed of hundreds of atoms, is to employ the so-called \textit{semiclassical approximation}. The core idea of this approximation is to neglect quantum correlations between atoms. In particular, one begins by writing down the Langevin-Heisenberg equations for the operators $\hat{\sigma}_{j}$, $\hat{\sigma}_{j}^{+}$, $\hat{\sigma}_{j}^{z}$, and then take the average values of both sides of the each equation. Solving the time-dependent differential equations for $\langle\hat{\sigma}_{j}\rangle$, $\langle\hat{\sigma}_{j}^{+}\rangle$, $\langle\hat{\sigma}_{j}^{z}\rangle$, however, requires knowing the time-dependent values of two-operator expectation values, such as $\langle\hat{\sigma}^{+}_{i}\hat{\sigma}_{j}\rangle$ and others, which, in turn, depend on three-operator expectation values and so on. This results in a cascaded hierarchy of a large number of differential equations. Within the semiclassical approximation, one assumes that the expectation value of a product of operators is always factorizable, i.e., $\langle\hat{A}\hat{B}\rangle \approx \langle\hat{A}\rangle \langle\hat{B}\rangle$ for any pair of operators $\hat{A}$ and $\hat{B}$~\cite{scully1999quantum, carmichael1999statistical}. This approximation allows to describe a system of $N$ atoms by solving a system of $3N$ coupled differential equations instead of $2^{2N}$.
In particular, the equations describing the evolution of the system are ~\cite{bettles_quantum_2020}:
\begin{align}\label{eq:res:06}
  \dot{\sigma}_{k} =& (-i \Delta_{k} - \frac{\Gamma_{kk}}{2}) \sigma_{k} + \sum_{j \neq k}\left(i \Omega_{jk} + \frac{\Gamma_{jk}}{2}\right) \sigma_{k}^{z}\sigma_{j} - i \frac{\mathbf{d}_{k} \mathbf{E}_{\text{in}}^{+}(\mathbf{r}_{k})}{\hbar} \sigma_{k}^{z} \\
  \dot{\sigma}_{k}^{+} =& (i \Delta_{k} - \frac{\Gamma_{kk}}{2}) \sigma_{k}^{+} + \sum_{j \neq k}\left(-i \Omega_{jk} + \frac{\Gamma_{jk}}{2}\right) \sigma_{j}^{+} \sigma_{k}^{z} + i \frac{\mathbf{d}_{k}^{*} \mathbf{E}_{\text{in}}^{-}(\mathbf{r}_{k})}{\hbar} \sigma_{k}^{z} \\
  \dot{\sigma}_{k}^{z} =& -\Gamma_{kk} \left(\sigma_{k}^{z} + 1 \right) - 2 \sum_{j \neq k}\left[ \left(i\Omega_{kj} + \frac{\Gamma_{kj}}{2} \right) \sigma_{k}^{+} \sigma_{j} + \left(-i \Omega_{jk} + \frac{\Gamma_{jk}}{2} \right) \sigma_{j}^{+}\sigma_{k}\right] \\
  &+ 2 i \left[\frac{\mathbf{d}_{k} \mathbf{E}_{\text{in}}^{+}(\mathbf{r}_{k})}{\hbar} \sigma_{k}^{+} - \frac{\mathbf{d}_{k}^{*} \mathbf{E}_{\text{in}}^{-}(\mathbf{r}_{k})}{\hbar} \sigma_{k}\right], \nonumber
\end{align}
where the first and second equations are the complex conjugated of each other.
The semiclassical approximation allows us to use the optimization procedure described above (based on maximizing $\mathcal{M}^{2}$, Eq.~\ref{eq:res:04}) for very large arrays. In Fig. \ref{fig:res:07} we show an example of an optimized geometry (parameters in caption) with $N_{\perp} = 10$ (total number of atoms $N=200$).

\begin{figure}[h!]
  \centering
  \begin{subfigure}[b]{0.4\textwidth}
    \includegraphics[width=\textwidth]{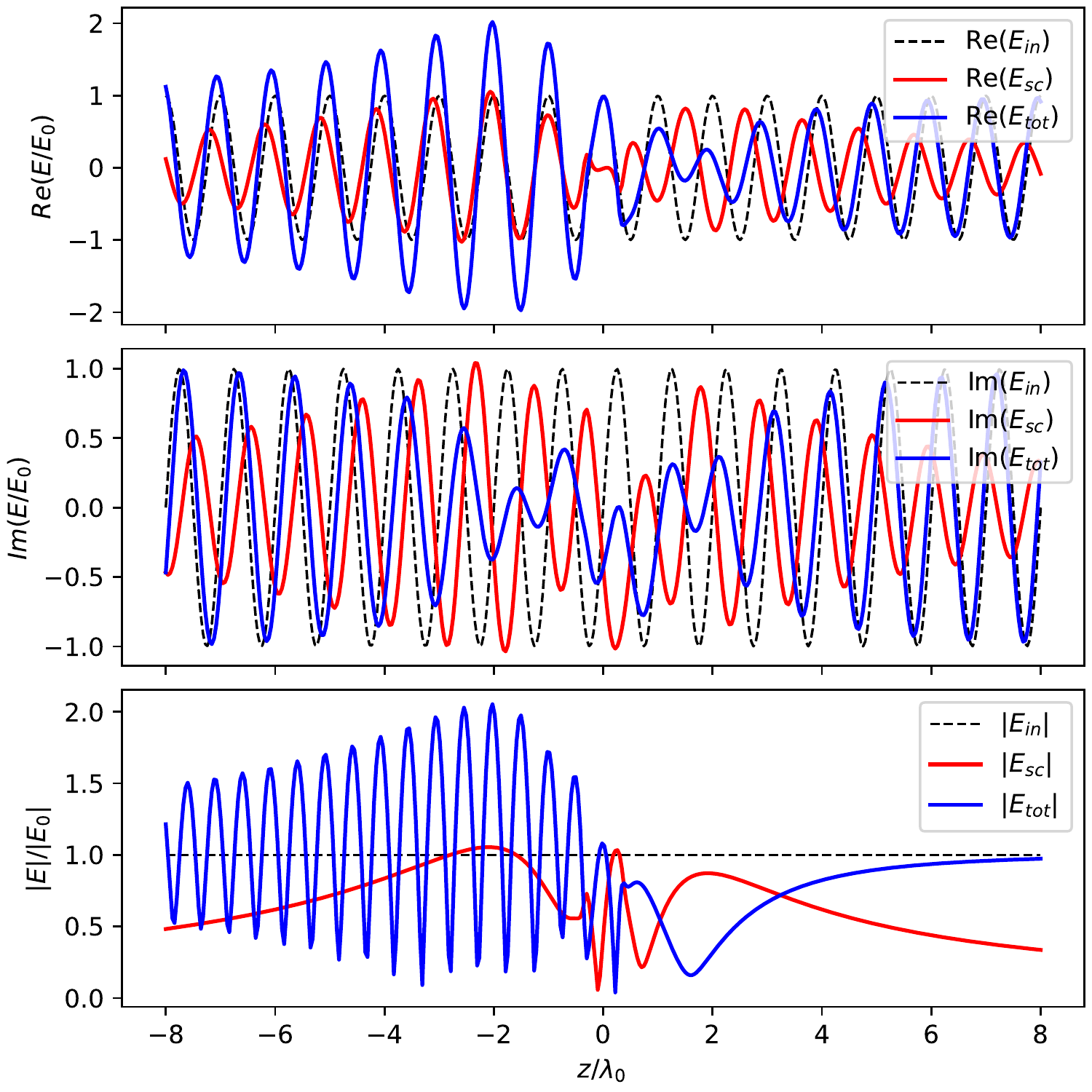}
    \caption{Forward direction.}
  \end{subfigure}
  \begin{subfigure}[b]{0.4\textwidth}
    \includegraphics[width=\textwidth]{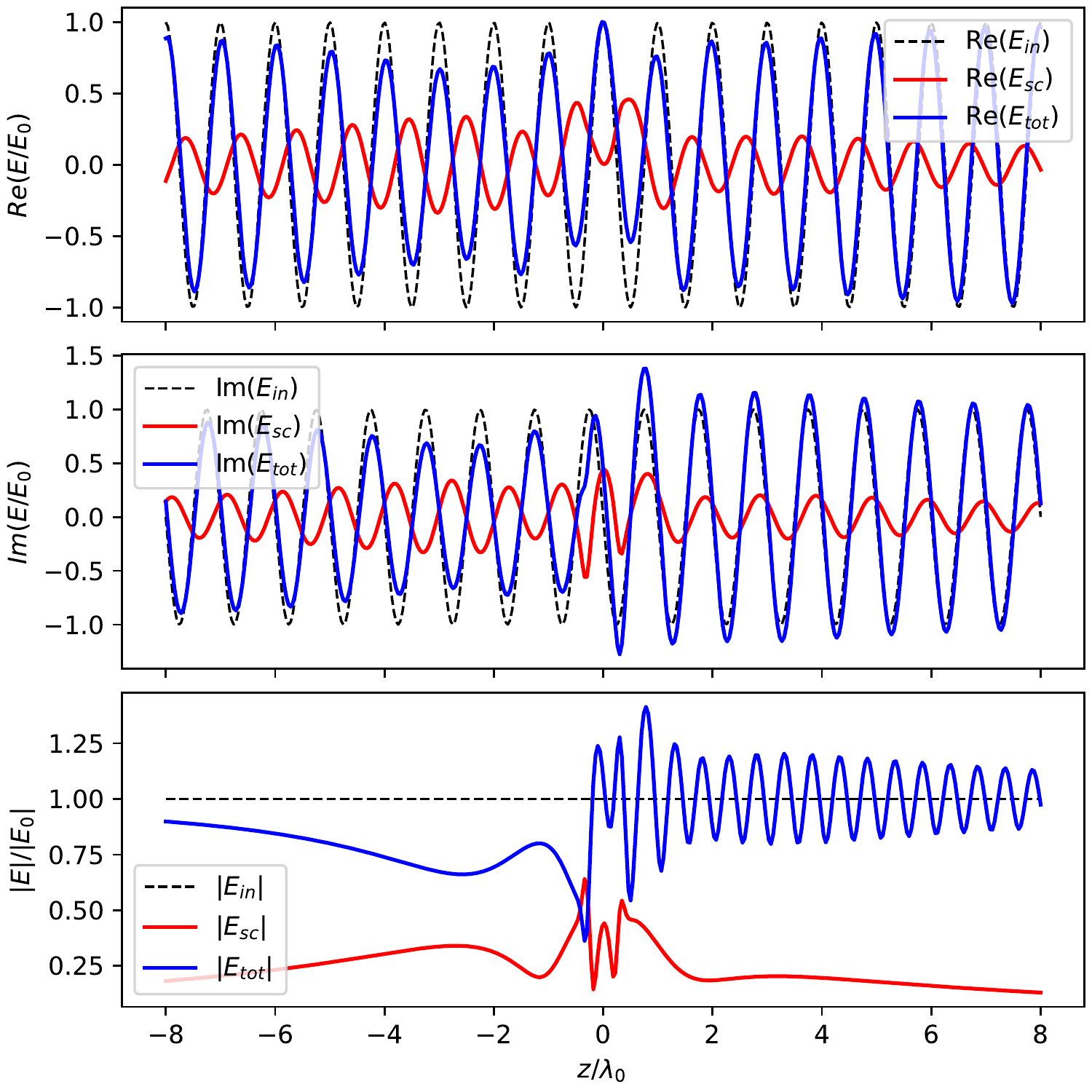}
    \caption{Backward direction.}
  \end{subfigure}
  \caption{Real part, imaginary part and absolute value of the $x$ component of incident, total and scattered fields at $x = 0$, $y = 0$, for the same system as in Fig.~\ref{fig:res:07}.}
  \label{fig:res:08}
\end{figure}

From Fig.~\ref{fig:res:07}, we see a similar behaviour of the scattered field as for the small arrays (compare with Fig.~\ref{fig:res:04}) -- the significant difference being the scattered field for the forward and backward directions of the excitation.
When excited along the forward direction, the scattered field (Fig.~\ref{fig:res:07}a, center panel) has a larger magnitude than when excited from the opposite direction.
The perturbation to the total field is also more noticeable in the case of the forward direction of excitation (Fig.~\ref{fig:res:07}a, left panel).
As mentioned earlier, the direction along which the scattering is larger is set by the phase shift between the arrays and the specific in-plane geometry.
For the particular set of optimal parameters considered here, differently from the small arrays considered earlier, for intermediate powers the system is trapped in the dark state when excited along the backward direction.

We can understand better the details of the scattering process by looking at 1D slices of the field distributions shown in Figs.~\ref{fig:res:07}.
Let us consider first the case of forward excitation direction (Fig.~\ref{fig:res:08}a): the scattered field (red solid lines) has a high amplitude, comparable to the one of the incident field ($E_{0}$, black dashed lines). For $z<0$ the real parts of the incident and scattered fields are in phase, while the imaginary parts cancel each other. This leads to the emergence of a standing-wave pattern of the total field for $z<0$ and to a large reflection level. For $z>0$ the incident and scattered fields interfere destructively, and the total field is suppressed for $\lambda_{0} < z < 3 \lambda_{0}$.
Beyond this region (i.e. for $z > 3\lambda_{0}$), the total field increases and its magnitude becomes comparable to the incident field. This is due to the fact that, since the incident field is assumed to be an infinitely extended plane wave, diffraction of the impinging field from the edges of the array dominates the total field at large distances, where the scattered field is instead almost zero.
For the backward excitation direction (Fig.~\ref{fig:res:08}b) the incident and scattered fields interfere destructively for $z<0$. However, due to the large difference between the amplitudes of these two fields, the total field is not fully canceled. For $z>0$ the incident and scattered fields have a mutual phase shift of $\approx \pi/2$, which does not lead to either destructive or constructive interference.

\begin{figure}[t!]
  \centering
  \includegraphics[width=0.6 \linewidth]{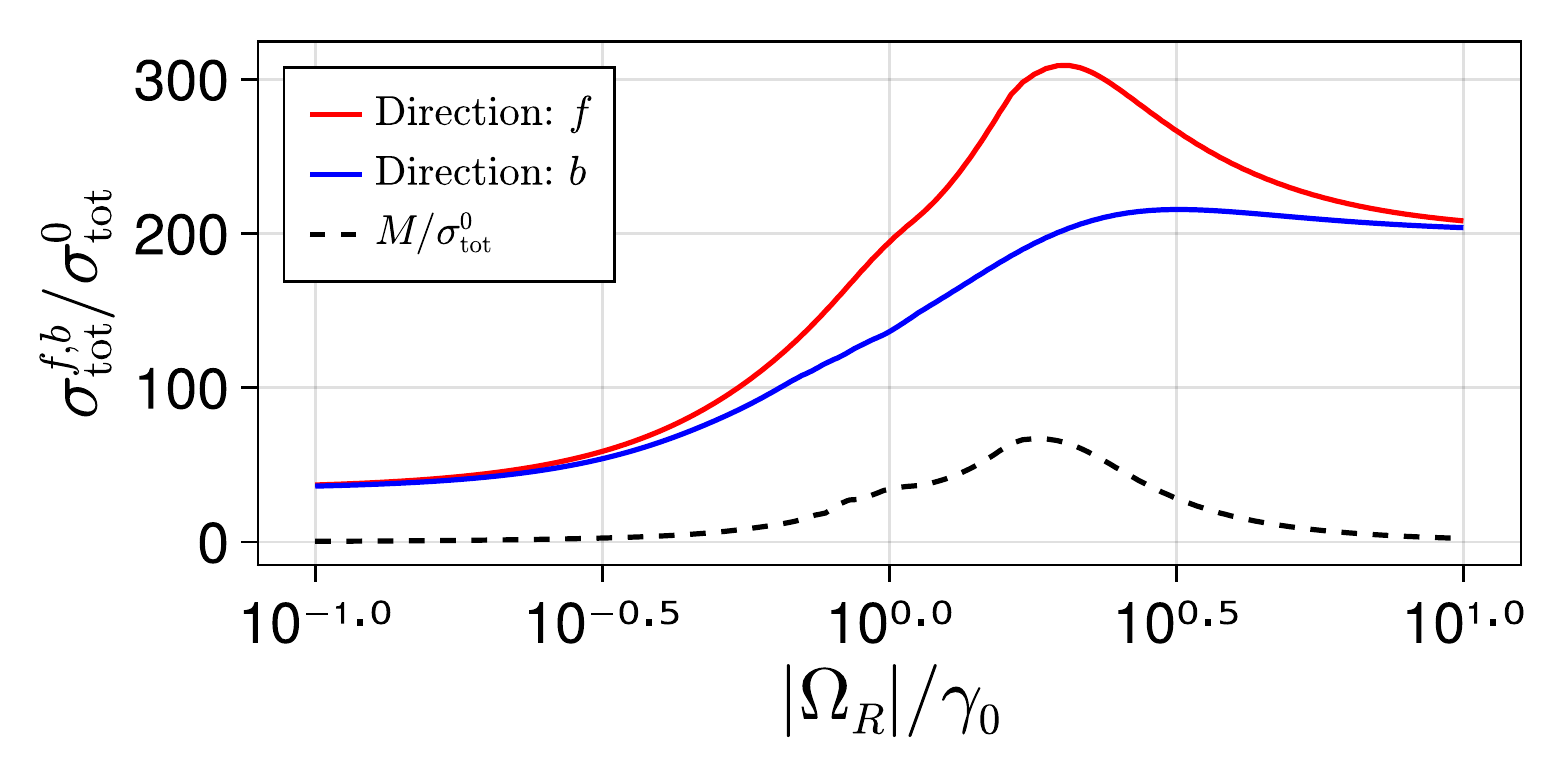}
  \caption{Total cross section associated with the forward and backward direction of excitation versus the absolute interaction constant between the incident field and the atoms, $|\Omega_{R}| / \gamma_{0}$, normalized by the total cross section of a single atom, $\sigma_{\mathrm{tot}}^{0}$, $N_{\perp} \times N_{\perp}= 100$. Parameters are the same as in Fig.~\ref{fig:res:07}.}
  \label{fig:res:09}
\end{figure}
The nonreciprocal extinction in this large system is also confirmed by the total cross sections. In Fig. \ref{fig:res:09} we plot the total cross sections for forward (red line) and backward (blue line) impinging directions versus the amplitude of the incident field. Similarly to what found in Fig.~\ref{fig:res:06}(a) for the case of small arrays, we found that  $\sigma_{\mathrm{tot}}^{f}$ and $\sigma_{\mathrm{tot}}^{b}$ are markedly different for intermediate powers.
Interestingly, we note that increasing the size of the system does not increase the nonreciprocal extinction efficiency $\mathcal{M}$. The maximum ratio between the two cross sections is $\sigma_{\mathrm{tot}}^{f} / \sigma_{\mathrm{tot}}^{b} \approx 1.5$, which is slightly smaller than the one obtained in Fig.~\ref{fig:res:06} for the two-dimer system. Moreover, we note that the maximum value of $\mathcal{M}$ occurs at higher values of input amplitude, $\Omega_{R} \approx 2 \gamma_{0}$.

\begin{figure}[h!]
  \centering
  \includegraphics[width=0.6 \linewidth]{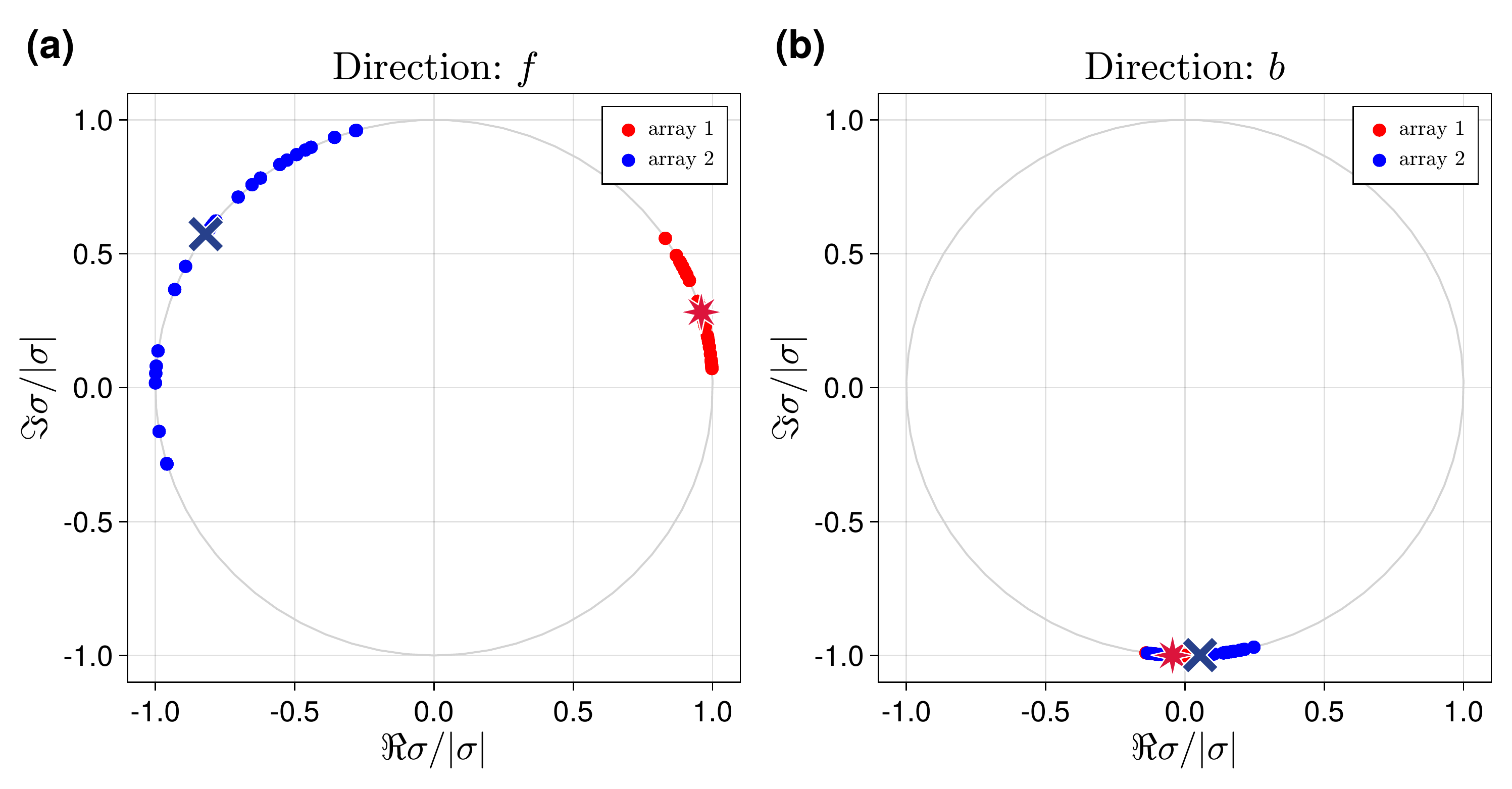}
  \caption{Phases of the atomic dipole moments in the steady state at optimal parameters for forward (panel a) and backward (panel b) direction of excitation. Red stars and blue crosses denote the average phases in the array 1 and array 2, respectively, $N_{\perp} \times N_{\perp} = 100$. Parameters are the same as in Fig.~\ref{fig:res:07}.}
  \label{fig:res:10}
  \centering
  \begin{subfigure}[b]{0.45\textwidth}
    \includegraphics[width=\textwidth]{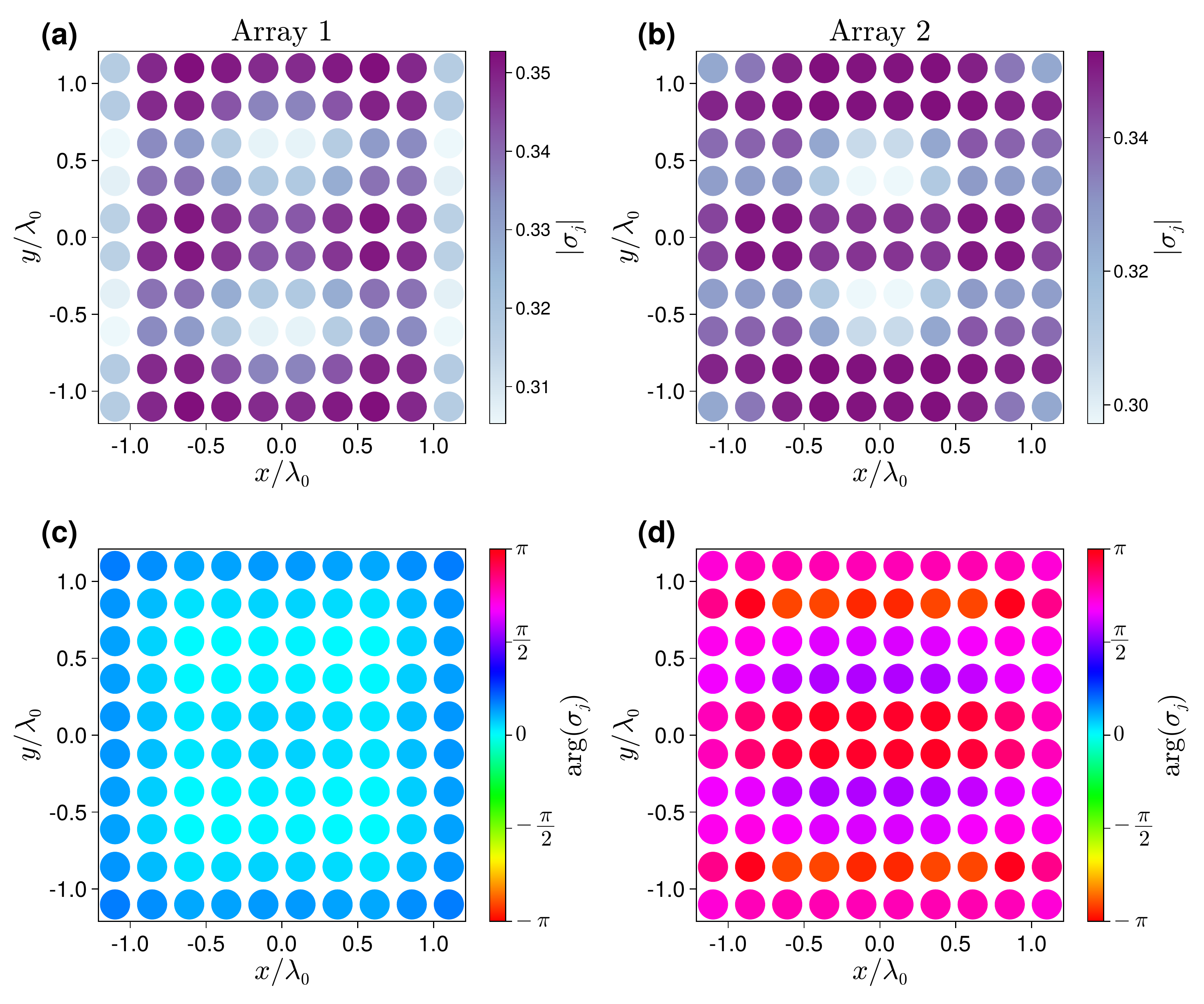}
    \caption{Forward excitation direction.}
  \end{subfigure}
  \begin{subfigure}[b]{0.45\textwidth}
    \includegraphics[width=\textwidth]{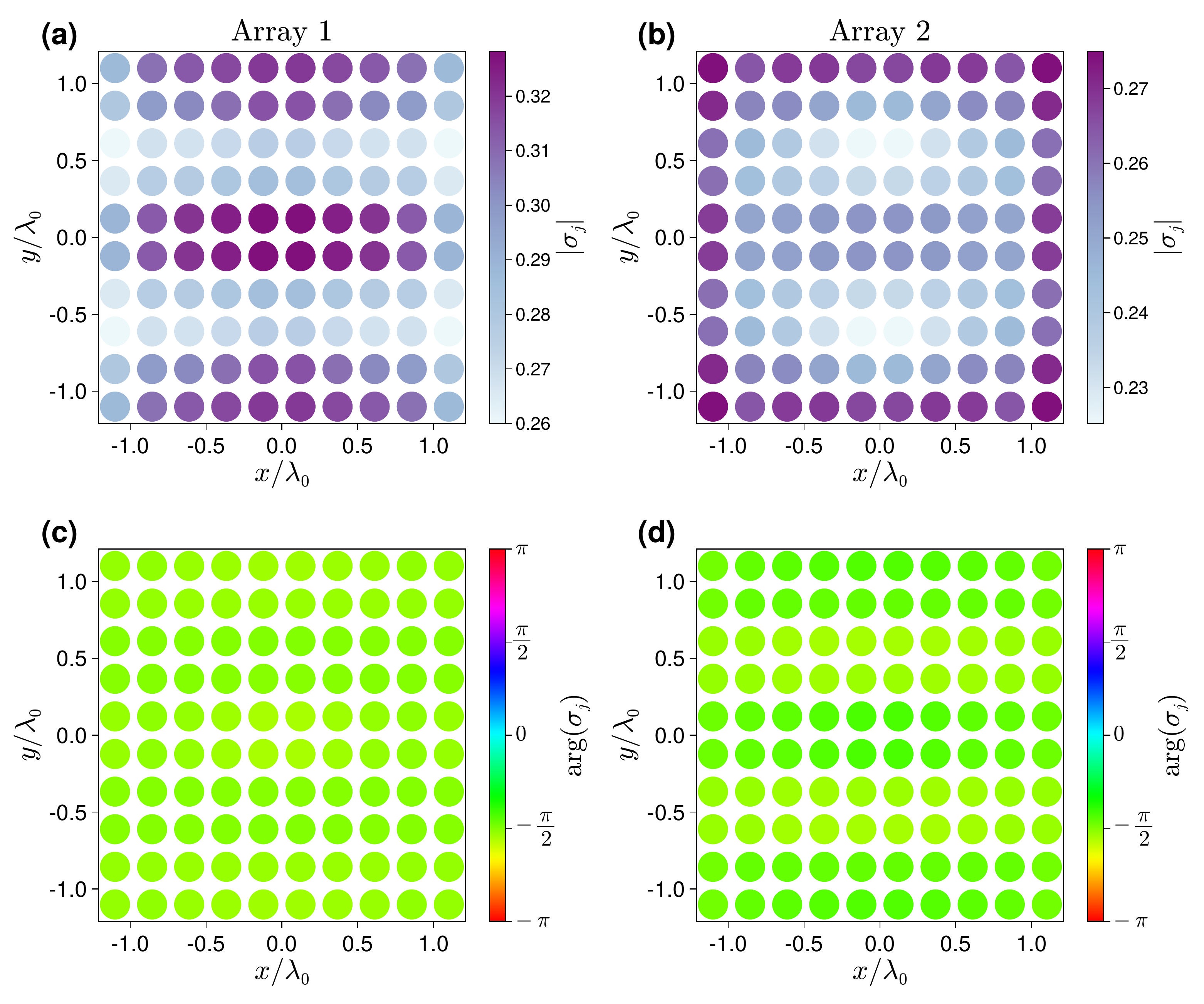}
    \caption{Backward excitation direction.}
  \end{subfigure}
  \caption{Absolute values and phases of the atomic dipole moments in the steady state at optimal parameters for forward (panel a) and backward (panel b) direction of excitation. $N_{\perp} \times N_{\perp} = 100$, other parameters are the same as in Fig.~\ref{fig:res:08}.}
  \label{fig:res:11}
\end{figure}

The connection between the nonreciprocal behaviour and the excitation of a dark state can be further proved by analyzing the distribution of the amplitudes and phases of the dipole moments of each atom (given by $\arg(\sigma_{j})$) in the steady state and at the power level where nonreciprocity is maximized. Figure~\ref{fig:res:10} shows the phases of the dipoles in polar plots, while Fig.~\ref{fig:res:11} shows the amplitude and phase of each atom in cartesian color-coded scatter plots.

For forward impinging direction (Fig.~\ref{fig:res:10}a), the phases of the dipoles are distributed over a large interval. On the contrary, for backward impinging direction (Fig.~\ref{fig:res:10}b) the phases are tightly localized in a small interval around $3\pi/2 $.
To quantify this phenomena, we define the average phase $\overline{\arg(\sigma)^d_{k}} = \sum_{j_{k}} \arg(\sigma^d_{j_{k}})/N^2_{\perp}$ and corresponding variance $\mathrm{Var}\left[\arg(\sigma)_{k}^{d}\right]$, where $j_{k}$ denotes the $j$-th atom of the $k$-th array ($k=1,2$), the subscript $\mathrm{d} = \mathrm{f},\mathrm{b}$ denotes the excitation direction, and the sum runs over all atoms in the array specified by $k$. For forward excitation direction,  the phases of the dipoles in array 1 are grouped around $\overline{\arg(\sigma)_{1}^{f}} \approx \pi / 10$ with a relatively small variance, $\mathrm{Var}\left[\arg(\sigma)_{1}^{f}\right] = 0.025$, whereas for array 2 we find $\overline{\arg(\sigma)_{2}^{f}} \approx 4 \pi / 5$, and the variance $\mathrm{Var}\left[\arg(\sigma)_{2}^{f}\right] = 0.25$ is larger by an order of magnitude. The difference between the average dipole phases in the two arrays is $|\overline{\arg(\sigma)_{1}^{0}} - \overline{\arg(\sigma)_{2}^{0}}| \approx 7 \pi / 10$, which is equals $\approx 17 \pi / 10$ when taking into account the phase shift between arrays ($\approx \pi$).

When considering the backward direction, instead, we observe a strong phasing of the dipole moments~\cite{nefedkin2016superradiance, nefedkin2017bad} both within each array and between array 1 and array 2.
The average phases are equal for both arrays, $\overline{\arg(\sigma)_{1}^{b}} \approx \overline{\arg(\sigma)_{2}^{b}} \approx - \pi/2$, and the variances are $\mathrm{Var}\left[\arg(\sigma)_{1}^{b}\right] = 2 \cdot 10^{-3}$ and $\mathrm{Var}\left[\arg(\sigma)_{2}^{b}\right] = 2 \cdot 10^{-2}$ for array 1 and array 2, respectively.
Combined with the fact that the arrays are separated by the distance $L \approx \lambda_{0} / 2$, this shows that the dipoles in the two arrays have almost the opposite phase. Opposite phases lead to suppression of the scattered field (see Fig.~\ref{fig:res:09}(b)), which is a manifestation of trapping the system's population in the dark state.
Importantly, despite the similar phases and their distributions across the arrays, the amplitudes of the atomic dipole moments are distributed differently on array 1 and array 2, see Fig.~\ref{fig:res:11}(b).

Thus, also within a semiclassical approximation approach, we found a clear connection between nonreciprocity and excitation of a dark state.
This trapping manifests itself as phasing of the dipoles in both arrays when the system is excited from one direction.
In addition, the dipoles in array 1 and array 2 have opposite phases, which also highlights the connection to a subradiant state.

\section{Conclusions}

In this work, we demonstrated large nonreciprocal extinction from asymmetric pairs of quantum metasurfaces composed of two-level atoms. Nonreciprocity is obtained by properly combining geometrical asymmetry and nonlinear response, hereby provided by the atoms saturable absorption. For a given number of atoms in the system, we optimized several parameters in the metasurface design to achieve the largest asymmetry between the total cross sections calculated when the system is excited from opposite directions. We have investigated systems composed of few atoms, which can be described by a fully-quantum master equation approach, and also systems composed of hundreds of atoms, whereby the semiclassical approximation is required to keep the numerical calculations feasible. In both cases, we showed that the occurrence of nonreciprocal extinction is intimately related to the asymmetric population of a dark state, and it results in nontrivial features in the spatial distribution of the scattered and total fields.
Moreover, we show that the nonreciprocal extinction, and the excitation of a dark state, also manifests peculiar phasing phenomena, whereby the phases of atomic dipoles condense near some common average phase within array 1 and array 2, and the phase shift between arrays is $\approx \pi$.

Our results show that the dark state plays a key role in the emergence of nonreciprocity in a system of distant atomic arrays, where atoms interact with each other through dipole-dipole interactions. We expect these results to stimulate experimental studies in atomic lattices of trapped cold atoms and quantum metasurfaces. Dark-state-induced nonreciprocity in atomic arrays may contribute to the development of quantum technologies that requiring efficient and tunable state transfer and state management at microscopic scale.

\begin{acknowledgments}
This work was supported by the Simons Foundation and the Air Force Office of Scientific Research.
\end{acknowledgments}

\appendix

\section{Derivation of the total cross section for one atom}

We consider a system consisting of a two-level atom, with transition frequency $\omega_a$ and dipole moment $\mathbf{d}$, located at $\mathbf{r}_0$ and excited by a classical monochromatic EM field with an arbitrary profile and frequency $\omega_0$.
The Hamiltonian of this system reads
\begin{equation} \label{eq:app:01}
\hat{H} = \hbar \omega_a \hat{\sigma}^+ \hat{\sigma} - \mathbf{d} \mathbf{E}^+_{\mathrm{in}}(\mathbf{r}_0) e^{-i \omega_0 t} \hat{\sigma} - \mathbf{d}^* \mathbf{E}^-_{\mathrm{in}}(\mathbf{r}_0) e^{i \omega_0 t} \hat{\sigma}^+.
\end{equation}
In order to get rid of the explicit time dependence, we rewrite the Hamiltonian in a frame rotating at the frequency of the incident field $\omega_0$.
The can be obtained via a  unitary transformation, described by the operator
\begin{align} \label{eq:app:02}
  \hat{U} &= \exp(-i \hbar \omega_0 \hat{\sigma}^+ \hat{\sigma})\\
  \hat{\bar{H}} &= \hat{U}^+ \hat{H} \hat{U} - i \hat{U}^+ \frac{\partial \hat{U}}{\partial t},
\end{align}
and the Hamiltonian in the rotating frame is
\begin{equation} \label{eq:app:03}
\hat{\bar{H}} = \hbar \Delta \hat{\sigma}^+ \hat{\sigma} - \hbar (\Omega_{R} \hat{\sigma} + \Omega_{R}^* \hat{\sigma}^+),
\end{equation}
where $\Delta = \omega_a - \omega_0$, and $\Omega_{R} = \mathbf{d} \mathbf{E}^+_{\mathrm{in}}(\mathbf{r}_0) / \hbar$. The atom interacts with the EM modes of free space and, therefore, spontaneously emits at the rate
\begin{equation} \label{eq:app_04}
\gamma_0 = \frac{\omega_0^3 d^2}{3\pi \epsilon_0 \hbar c^3}.
\end{equation}
After applying Born and Markov approximations, we obtain the master equation in Lindblad form~\cite{carmichael1999statistical},
\begin{align}
\dot{\hat{\rho}} =& -\frac{i}{\hbar} \left[\hat{\bar{H}}, \hat{\rho} \right] \nonumber \\
& + \frac{\gamma_0}{2} \left(1 + n(\omega_a, T)\right) (2 \hat{\sigma} \rho \hat{\sigma}^+ - \hat{\rho} \hat{\sigma}^+ \hat{\sigma} - \hat{\sigma}^+ \hat{\sigma} \hat{\rho}) \nonumber \\
& + \frac{\gamma_0}{2} n(\omega_a, T) (2 \hat{\sigma}^+ \hat{\rho} \hat{\sigma} - \hat{\rho} \hat{\sigma} \hat{\sigma}^+ - \hat{\sigma} \hat{\sigma}^+ \hat{\rho}), \nonumber
\end{align}
where
$n(\omega_a, T) = 1 / \left(e^{\hbar \omega_a / k_B T} - 1\right)$ is the Bose-Einstein distribution. By looking for the steady-state of the master equation, we find the steady-state density matrix,
\begin{align}
	\rho_{11}^\text{s.s.} =& \frac{\frac{4 |\Omega_{R}|^2}{\gamma _0^2 \left(2 n\left(\omega _a,T\right)+1\right){}^2+8 |\Omega_{R}|^2+4 \Delta ^2}+n\left(\omega _a,T\right)}{2 n\left(\omega _a,T\right)+1}, \label{eq:ex:AF:1} \\
	\rho_{12}^\text{s.s.} =& \frac{2 i \Omega_{R} ^* \left(\gamma _0 \left(2 n\left(\omega _a,T\right)+1\right)+2 i \Delta \right)}{\left(2 n\left(\omega _a,T\right)+1\right) \left(\gamma _0^2 \left(2 n\left(\omega _a,T\right)+1\right){}^2+8 |\Omega_{R}|^2+4 \Delta ^2\right)}, \label{eq:ex:AF:2}\\
	\rho_{21}^\text{s.s.} =& -\frac{2 \Omega_{R}  \left(2 \Delta +i \gamma _0 \left(2 n\left(\omega _a,T\right)+1\right)\right)}{\left(2 n\left(\omega _a,T\right)+1\right) \left(\gamma _0^2 \left(2 n\left(\omega _a,T\right)+1\right){}^2+8 |\Omega_{R}|^2+4 \Delta ^2\right)}, \label{eq:ex:AF:3}\\
	\rho_{22}^\text{s.s.} =& 1 - \frac{\frac{4 |\Omega_{R}|^2}{\gamma _0^2 \left(2 n\left(\omega _a,T\right)+1\right){}^2+8 |\Omega_{R}|^2+4 \Delta ^2}+n\left(\omega _a,T\right)}{2 n\left(\omega _a,T\right)+1}. \label{eq:ex:AF:4}
\end{align}
Following standard procedures, we can calculate the average value of the atom polarization as $\langle \hat{\mathbf{p}} \rangle = \mathrm{Tr}(\hat{\rho}^{\mathrm{s.s.}} \hat{\mathbf{p}})$ assuming that $n(\omega_a, T) \approx 0$, resulting into
\begin{equation} \label{eq:res:04_1}
	\langle \hat{\mathbf{p}} \rangle = \mathbf{d} \langle \hat{\sigma} \rangle = \mathbf{d} \frac{2 i (\gamma_{0} - 2 i \Delta ) \Omega_{R}^*}{\gamma_{0}^2+8 |\Omega_{R}|^2+4 \Delta ^2}.
\end{equation}
We then substitute the average polarization $\langle \hat{\mathbf{p}} \rangle$ into Eq.~(\ref{eq:res:02}) and using Eq.~(\ref{eq:res:03}) we find the total cross section of 1 atom, $\sigma_{\mathrm{tot}}^{\mathrm{a}}$.

\section{Further analysis of the dimer system}

\subsection{Spectral properties of the system}
In the main text, the frequency of the incident field was fixed to a certain value $\omega_{0}$ and the system geometry and parameters where then optimized to maximize nonreciprocity. In this section we discuss the dependence of the total cross section, the von Neumann entropy of the system and the dark state population on the input frequency $\omega_{0}$, for the same parameters considered in the main text (see Sec.~\ref{subsec:quantum} and the caption of Fig.~\ref{fig:res:05}).
In the main text the atoms were symmetrically detuned with respect to the incident EM field frequency, i.e., the input frequency was fixed to $\omega_{0} = \omega_{1,2} \equiv (\omega_1 + \omega_2)/2$.

\begin{figure}[h!]
  \centering
  \includegraphics[width=0.45\linewidth]{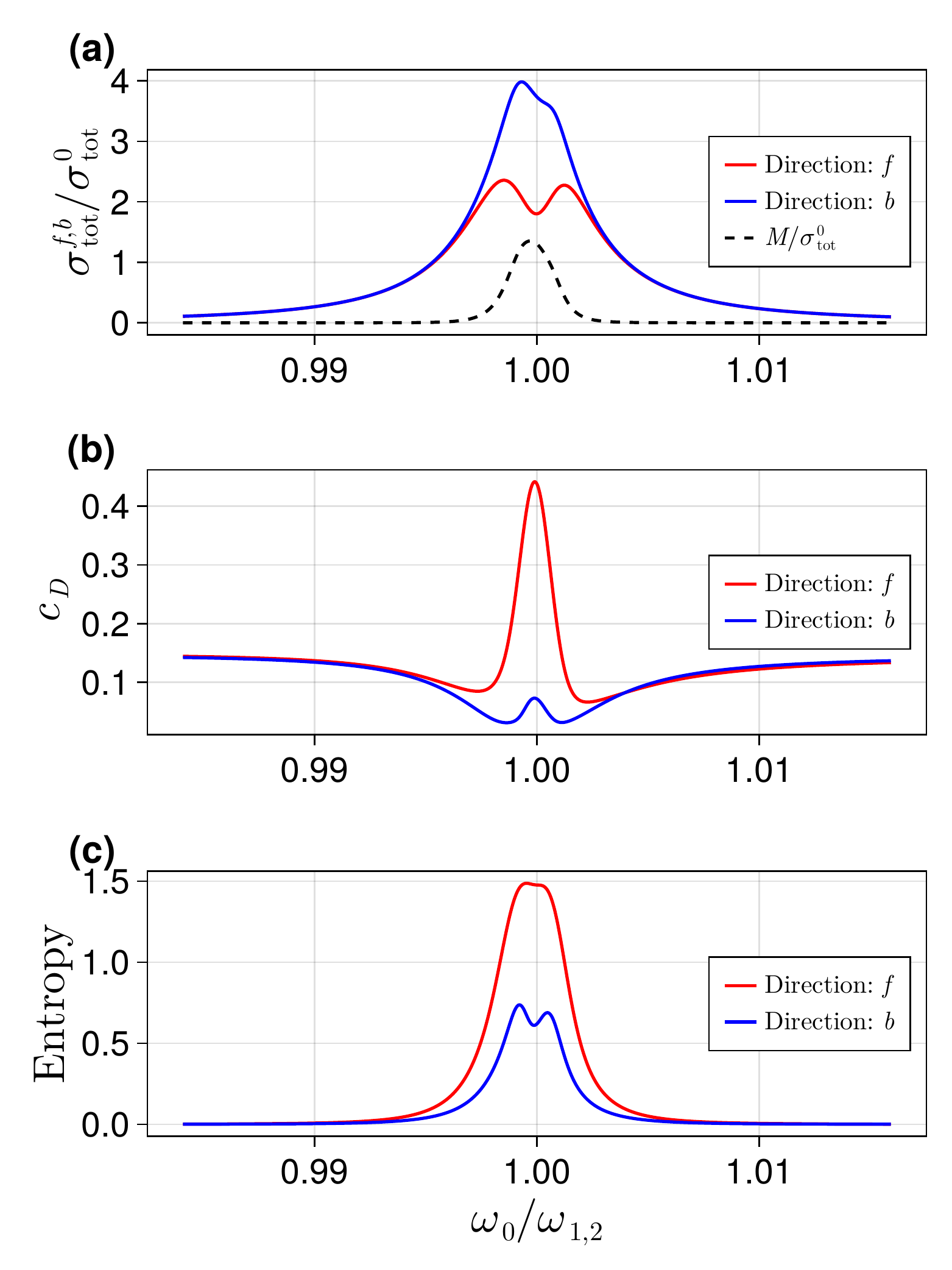}
  \caption{(a) Total cross section and nonreciprocal extinction efficiency, (b) population of the dark state, and (c) von Neumann entropy of the two-dimer system (with total number of atoms $N = 4$) when exciting along the forward direction (red line) and the backward direction (blue line), versus the frequency of the incident field. The black dashed line in panel (a) shows the corresponding value of the  nonreciprocal extinction efficiency $\mathcal{M}$. The system parameters are $\Delta = -\gamma_{0}$, $\delta = a_{2} - a_{1} = \lambda_{0} / 10$, $a_{1} = \lambda_{0}/3$, $L = \lambda_{0} / 10$.}
  \label{fig:app:00}
\end{figure}

Fig.~\ref{fig:app:00} shows the dependence of the total cross sections, dark state populations and von Neumann entropy for the forward and backward directions of excitation on the frequency of the incident field (normalized by the average atomic frequency, $\omega_{1,2}$).
The total cross sections have a spectral bandwidth determined by the dark state decay rate, $\gamma_{D}$ and, therefore, the nonreciprocity effect is observable within a comparable bandwidth $\Delta \omega \approx \gamma_{D}$,.
At the same time, due to the optimization procedure described in the main text, the maximum of $\mathcal{M}$ is reached when the incident field frequency is approximately equal to the average atomic frequency, $\omega_{0} = \omega_{1,2}$.
Moreover, both the population of the dark state, $c_{D}$, and the von Neumann entropy have the maximum contrast for forward and backward directions of excitation at $\omega_{0} = \omega_{1,2}$ as well.
The doublet of peaks visible in some of the panels of Fig. \ref{fig:app:00} is due to the detuning between atoms in array 1 and array 2.

\subsection{Total and scattered fields}
\begin{figure}[h!]
  \centering
  \begin{subfigure}{0.8\textwidth}
    \includegraphics[width=0.9\linewidth]{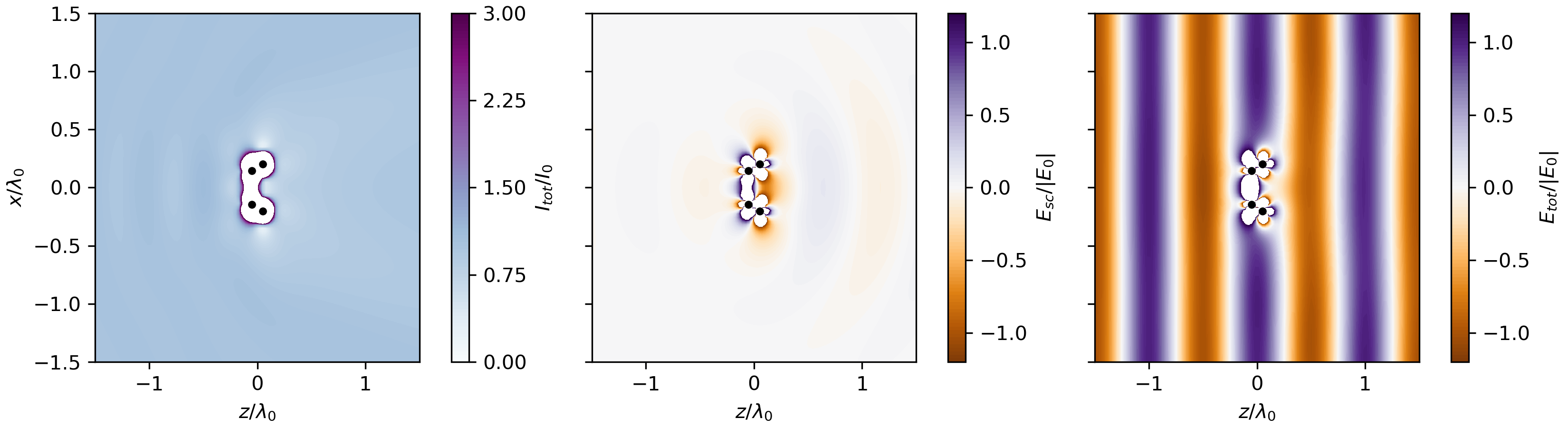}
    \caption{Forward direction.}
  \end{subfigure}
  \vfill
  \begin{subfigure}{0.8\textwidth}
    \includegraphics[width=0.9\linewidth]{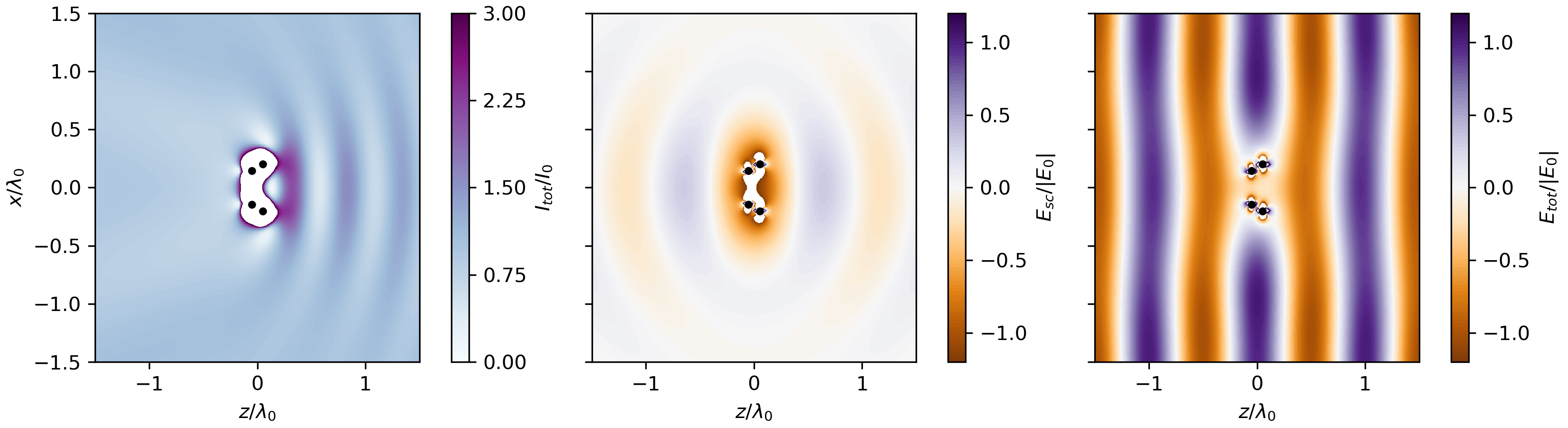}
    \caption{Backward direction.}
  \end{subfigure}
  \caption{Total field intensity, real part of $E_\text{sc}^{x}$ and real part of $E_\text{tot}^{x}$ upon forward (panel a) and backward (panel b) excitation of the two-dimer system, each of them composed by 2 atoms. Parameters: $\Delta = -\gamma_{0}$, $\delta = a_{2} - a_{1} = \lambda_{0} / 10$, $a_{1} = \lambda_{0}/3$, $L = \lambda_{0} / 10$.}
  \label{fig:app:01}
\end{figure}

In the main text we studied nonreciprocity emerging in the simple dimer system (whose geometry is sketched in Fig.~\ref{fig:res:06}(c)).
Using the approach developed in the main text, we found the total field and the scattered field for the dimer system at the optimal parameters which maximize the nonreciprocal extinction efficiency (\ref{eq:res:04}), see the caption of Fig.~\ref{fig:res:05}.
In Fig.~\ref{fig:app:01} we show additional results for this system, namely the total intensity, scattered field and total field.
From Fig.~\ref{fig:app:01}, one can see that all the characteristic features of the fields behaviour matches the case of a larger arrays shown in Fig.~\ref{fig:res:07}.

\subsection{Steady-state density matrix at the high power}
\begin{figure}[h!]
  \centering
  \includegraphics[width=0.33 \linewidth]{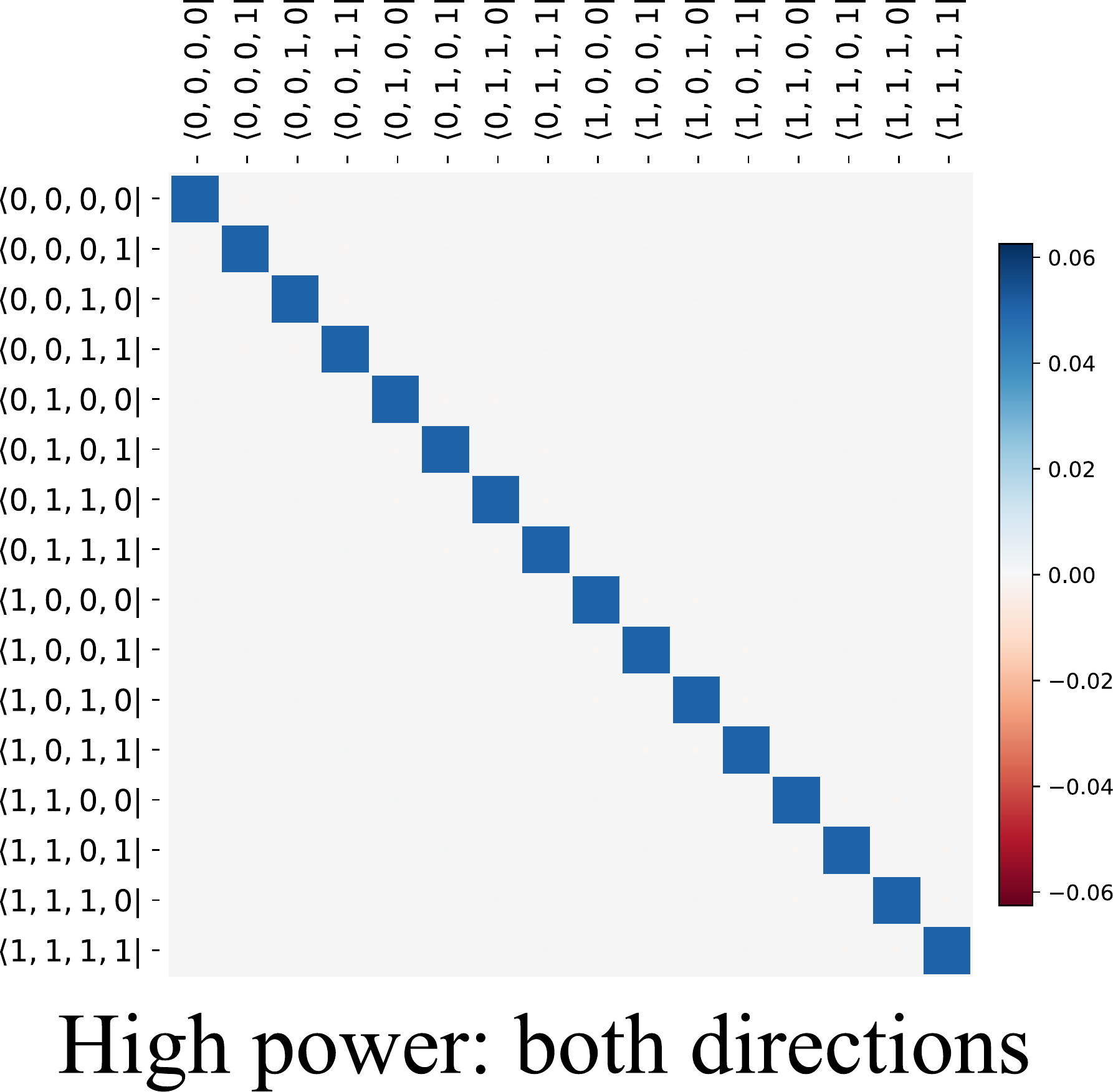}
  \caption{The elements of the steady-state density matrices when the two-dimer system is excited along either the forward and backward direction. The density matrix is plotted in the basis of the uncoupled atomic states. The system parameters are $\Delta = -\gamma_{0}$, $\delta = a_{2} - a_{1} = \lambda_{0} / 10$, $a_{1} = \lambda_{0}/3$, $L = \lambda_{0} / 10$, $|\Omega_R| = 20 \gamma_0$.}
  \label{fig:app:02}
\end{figure}

Further let us have a look at the steady state of the system at high powers of excitation, i.e., when $|\Omega_{R}| \gg \gamma_{0}$.
The corresponding density matrix is shown in Fig.~\ref{fig:app:02}.
Here the density matrix is close to a diagonal form, i.e. all the nondiagonal elements are negligible.
Moreover, the diagonal elements which are responsible for the populations of the correspondent states are all equal.
This fact is reflected in the behaviour of the von Neumann entropy, which reaches its maximum at the high power of the incident field, see Fig.~\ref{fig:res:06}(d).

\bibliography{ref_AA}

\end{document}